\documentclass{jpp}
\usepackage{graphicx}

\usepackage[utf8]{inputenc}
\usepackage[T1]{fontenc}
\usepackage{amsmath}
\usepackage{xcolor}
\usepackage{subcaption}
\usepackage{siunitx}
\usepackage{booktabs}
\usepackage{float}

\newcommand{\pderiv}[2]{
\frac{\partial #1}{\partial #2}
}

\newcommand{\gyavgR}[1] {
  \langle{ #1 \rangle}_{\boldsymbol {R}}
}

\newcommand{\gyavgr}[1] {
  \langle{ #1 \rangle}_{\boldsymbol{r}}
}

\newcommand{\vEM} {
  {\boldsymbol{v}}_{\chi}
}

\newcommand{\bhat}{
  {\boldsymbol{\hat{b}}}
}

\shorttitle{Implicit-Explicit Methods for the GX Gyrokinetics Code}
\shortauthor{P. S. Kim et al.}

\title{Implicit-Explicit Methods for the {\tt GX} Gyrokinetics Code}

\author{P. S. Kim\aff{1, 2}
  \corresp{\email{pk2354@princeton.edu}},
  N. R. Mandell\aff{2, 3},
  I. G. Abel\aff{4},
  W. Dorland\aff{4, 5},
  G. W. Hammett\aff{1, 2},
  \and D. R. Reynolds\aff{6}
  }

\affiliation{\aff{1}Princeton University, Princeton, NJ 08544, USA,
\aff{2}Princeton Plasma Physics Laboratory, PO Box 541, Princeton NJ 08543, USA,
\aff{3} Type One Energy, Knoxville, TN 37931, USA
\aff{4} Institute for Research in Electronics and Applied Physics, University of Maryland, College Park, MD 20742, USA,
\aff{5} University of Maryland, College Park, MD 20742, USA,
\aff{6} Department of Mathematics \& Statistics, University of Maryland, Baltimore County, Baltimore, MD 21250, USA
}

\begin{document}

\maketitle

\begin{abstract}
Gyrokinetic simulations are crucial for understanding turbulent transport in magnetic confinement fusion devices. Unfortunately, these simulations are typically extremely expensive, particularly when kinetically evolving electrons due to their significantly greater thermal speed compared to that of ions. In this work, we present new implicit-explicit timestepping methods for the {\tt GX} gyrokinetics code to accelerate these simulations where the ion and electron streaming terms as well as the electron mirror terms  are treated implicitly, while the remaining terms are treated explicitly. These methods include a lower-order Lie-Trotter splitting method as well as a higher-order additive Runge-Kutta method. Our schemes take advantage of the sparsity of the GX equations, which allows us to efficiently invert the linear operators that arise from the implicit scheme without consuming significant amounts of limited GPU memory. We demonstrate our new integration scheme for both linear (electrostatic and electromagnetic) and nonlinear (electrostatic) simulations in general toroidal geometry, and show that our methods can achieve up to the maximum theoretical linearly stable timestep for these simulations while maintaining a significant improvement in wall-clock time.\end{abstract}

\section{Introduction}
Tokamaks and stellarators are two of the leading nuclear fusion concepts, and use strong magnetic fields to confine the fusion plasma. Assuming that the plasma is macroscopically stable, tokamaks and stellarators still leak energy and particles due to various transport processes \citep{Hinton1976TheorySystems, Horton1999DriftTransport}. The largest contributor to these losses is often turbulent transport. Driven by steep temperature and density gradients, turbulent fluctuations rapidly mix the hot and cold regions of the plasma, preventing the plasma from maintaining higher pressures and limiting the overall fusion power gain \citep{Liewer1985MeasurementsTransport, Wootton1990FluctuationsTokamaks, Dimits2000ComparisonsSimulations}.

One of the most widely-used models for simulating magnetized turbulence is gyrokinetics \citep{AntonsenJr.1980KineticPlasmas, Frieman1982NonlinearEquilibria, Abel2013MultiscaleFlows, Catto1978LinearizedGyro-kinetics}. Numerous gyrokinetic codes \citep{Kotschenreuther1995ComparisonInstabilities, Dorland2000ElectronTurbulence, Jenko2000ElectronTurbulence, Candy2003AnSolver, Barnes2019Stella:Configurations, Mandell2024GX:Design} have been developed in the past few decades that can accurately match experimental measurements of turbulent losses as well as temperature and density profiles. Unfortunately, running these gyrokinetic codes can be computationally expensive, often requiring hundreds to thousands of CPU or GPU-hours. Including kinetic electrons exacerbates this cost, as their higher thermal speed (about 60 times greater than a deuterium ion of the same temperature) enforces a stricter Courant–Friedrichs–Lewy (CFL) condition that reduces the maximum allowed timestep. To avoid this issue, an adiabatic electron response is often used, particularly when studying the ion-temperature gradient mode \citep{Coppi1967InstabilitiesConfigurations, Cowley1991ConsiderationsTurbulence}. However, correctly resolving the electron dynamics is crucial for understanding other important microinstabilities like the trapped-electron mode \citep{Rosenbluth1968LowFrequencyInstability}, the kinetic-ballooning mode \citep{Tang1980Kinetic-ballooning-modeGeometry}, and the micro-tearing mode \citep{Hazeltine1975KineticInstability, Guttenfelder2011ElectromagneticTurbulence}.

To overcome this computational restriction, codes like {\tt GS2} \citep{Kotschenreuther1995QuantitativeEffects, Dorland2000ElectronTurbulence} and {\tt stella} \citep{Barnes2019Stella:Configurations} have employed implicit or implicit-explicit (IMEX) time-integration methods. {\tt GS2} is fully linearly implicit, while {\tt stella} treats the stiff electron terms implicitly. Here, linearly implicit means that only the linear part of the gyrokinetic equation is treated implicitly, while the nonlinear term appears explicitly and can be evaluated separately through operator splitting. Among the linear terms, only a subset of them are stiff due to the higher electron thermal speed. Alternatively, in {\tt CGYRO} \citep{Candy2016APlasmas}, one of the velocity coordinates is truncated, relaxing the CFL constraint. In this paper, we focus on the {\tt GX} gyrokinetic code \citep{Mandell2018Laguerre-HermiteGyrokinetics, Mandell2024GX:Design}. {\tt GX} is a GPU-native code that solves the radially-local $\delta f$ form of gyrokinetics by expanding the distribution function in a Hermite-Laguerre basis in velocity space,  allowing for coarser velocity resolution. For our scheme, we treat implicitly the terms related to parallel streaming along field lines for both ions and electrons, as well as the term due to acceleration due to the mirror force for electrons.

Although being GPU-native substantially improves performance, the limited memory on GPUs can be very constraining for implicit methods. Inverting the linear operators that arise in implicit methods often requires significant storage. Indeed, as we explain in \S\ref{k's method}, the memory cost of the implicit methods implemented in {\tt GS2} and {\tt stella} can make them nearly infeasible to use for the GPU-native {\tt GX}. However, the pseudo-spectral formulation used in {\tt GX} yields equations that are sparse in velocity-space. We take advantage of this sparsity to present new algorithms to efficiently invert the implicit operators. We also employ physics-based preconditioners and iterative methods to further reduce the memory cost. We note that a similar Hermite–Laguerre code, {\tt GKX}, has been recently developed that also utilizes IMEX timestepping \citep{Jorge2026GKX:JAX}, but it inverts the implicit operator exactly, using a direct factorization that is significantly more memory expensive than our low-memory approximate inversions.

The paper is organized as follows: In  \S\ref{gx intro}, we introduce the {\tt GX} code, and the resulting equations that need to be solved. Section \ref{time integration schemes} describes the time-integration schemes that we use in this work. Our methods to invert the implicit operators are described in  \S\ref{advance implicit}, and simulation examples are shown in \S\ref{numerical results}. The conclusions follow in \S\ref{conclusions}.

\section{The {\tt GX} Gyrokinetics Code}\label{gx intro}
Our notation follows that of \citet{Mandell2018Laguerre-HermiteGyrokinetics, Mandell2024GX:Design}. {\tt GX} solves the radially local $\delta f$ formulation of gyrokinetics. In this approximation, the distribution function for species $s$ is split as $F_s = F_{0s} + \delta f_s$, where $\delta f_s$ is a perturbation to an equilibrium distribution function $F_{0s}$ that is assumed to be Maxwellian:
% \begin{equation}
%     F_{0s} = F_{Ms} = \frac{n_s}{\left(2\upi v_{ts}^2\right)^{3/2}}\mathrm{e}^{-v_{\parallel}^2/2 - \mu B}.
% \end{equation}

\begin{equation}
    F_{0s} = F_{Ms} = \frac{n_s}{\left(2\upi v_{th,s}^2\right)^{3/2}}
    \exp\left(-\frac{v_\parallel^2/2 + \mu B}{v_{th,s}^2}\right),
\end{equation}
where $n_s$ is the equilibrium density of species $s$, the thermal speed is $v_{th,s} = \sqrt{T_s/M_s}$, and $T_s$ and $M_s$ are the equilibrium temperature and mass of species $s$, respectively.

{\tt GX} uses as velocity-space coordinates the velocity parallel to the background magnetic field $v_{\parallel}$ and $\mu B$. The magnetic moment $\mu$ is $\mu = v_{\perp}^2/2B$, where $v_{\perp}$ is the speed perpendicular to the magnetic field, and $B$ is the magnitude of the local magnetic field. Such velocity-space coordinates are useful for complex stellarator geometries that may have several magnetic wells along a field line. The perturbation is further split into non-Boltzmann and Boltzmann parts 

\begin{equation}
    \delta f_s = h_s - \frac{q_s \Phi}{T_s} F_{Ms}
    = g_s + \frac{q_s \gyavgR{\chi}}{T_s} F_{Ms} - \frac{q_s \Phi}{T_s} F_{Ms},
\end{equation}
where $q_s$ is the charge of species $s$, $\chi = \Phi - \boldsymbol{v}\bcdot\boldsymbol{A}/c$ with $c$ the speed of light (both {\tt GX} and this paper are written in Gaussian units), and $\Phi$ and $\boldsymbol{A}$ are the electrostatic and magnetic vector potential, respectively.

The non-Boltzmann part $h_s$ and the auxiliary distribution function $g_s=h_s - q_s/T_s \gyavgR{\chi} F_{Ms}$ satisfy the gyrokinetic equation
% \begin{align}
% \pderiv{g_s}{t} + &\left[v_{ts}  v_\parallel \bhat + \gyavgR{\vEM} +
%                   \frac{\tau_s}{Z_s}{\bf v}_d \right] \bcdot \bnabla h_s - v_{ts} \mu \left(\bhat\bcdot\bnabla B\right) \pderiv{h_s}{v_\parallel} \\ 
%                   &= - \gyavgR{\vEM} \bcdot \bnabla\big|_E F_{Ms} + C(h_s).
% \label{gkg}
% \end{align}
% \begin{align}
% \pderiv{g_s}{t} + &\left[v_\parallel \bhat + \gyavgR{\vEM} +
%                   {\bf v}_{ds} \right] \bcdot \bnabla h_s - \mu \left(\bhat\bcdot\bnabla B\right) \pderiv{h_s}{v_\parallel} \notag \\ 
%                   &= - \gyavgR{\vEM} \bcdot \bnabla\big|_E F_{Ms} + C(h_s),
% \end{align}
\begin{equation}
\begin{split}
\pderiv{g_s}{t} + &\left[v_\parallel \bhat + \gyavgR{\vEM} +
                  {\boldsymbol v}_{ds} \right] \bcdot \bnabla h_s
                  - \mu \left(\bhat\bcdot\bnabla B\right) \pderiv{h_s}{v_\parallel} \\
                  &= - \gyavgR{\vEM} \bcdot \bnabla\big|_E F_{Ms} + C(h_s),
\end{split}
\end{equation}
where $\gyavgR{\vEM} = c\,\bhat \times \bnabla \gyavgR{\chi}/B$ is the gyroaveraged fluctuating drift velocity, ${\boldsymbol v}_{ds} = \bhat \times \left( v_\parallel^2\, \bhat\bcdot\bnabla\bhat + \mu \bnabla B \right)/\Omega_s$ contains the magnetic curvature and $\bnabla B$ drifts, and $\Omega_s = q_s B/(M_s c)$ is the cyclotron frequency. The collision operator is represented by $C(h_s)$.
% \begin{equation}
%     \gyavgR{\vEM} = \frac{\bhat \times \bnabla \gyavgR{\chi}}{B}, \qquad
%     {\bf v}_{ds} = \frac{\bhat}{\Omega_s} \times \left( v_\parallel^2\, \bhat\bcdot\bnabla\bhat + \mu \bnabla B \right),
% \end{equation}
% with $\Omega_s = q_s B/M_s$, $\mu = v_\perp^2/2B$, and $\chi = \Phi - \boldsymbol{v}\bcdot\boldsymbol{A}$.

The auxiliary distribution function $g_s$ was used to eliminate the time derivative of the gyroaveraged fields. These fields are self-consistently determined using the following field equations
% \begin{gather}
%     \sum_s \frac{ Z_s^2 n_s}{\tau_s} \Phi = \sum_s Z_s n_s \int d^3{\bf v}\ \gyavgr{h_s}, \label{qneuts} \\
%     -\bnabla_\perp^2 A_\parallel =\frac{\beta_\mathrm{ref}}{2}\sum_s Z_s n_s v_{ts}  \int d^3{\bf v}\ v_\parallel  \gyavgr{h_s}, \label{ampere} \\
%     \bnabla_\perp^2 \delta\! B_\parallel = -\frac{\beta_\mathrm{ref}}{2 {\color{black} B}}\bnabla_\perp\bnabla_\perp : \sum_s n_s \tau_s \int d^3\vec{v}\ \langle \vec{v}_\perp \vec{v}_\perp h_s\rangle_r, \label{dB}
% \end{gather}
\begin{gather}
    \sum_s \frac{q_s^2 n_s}{T_s} \Phi = \sum_s q_s \int \mathrm{d}^3{\boldsymbol v}\ \gyavgr{h_s}, \\
    -\bnabla_\perp^2 A_\parallel = \frac{4\upi}{c} \sum_s q_s \int \mathrm{d}^3{\boldsymbol v}\ v_\parallel \gyavgr{h_s}, 
\end{gather}
and
\begin{equation}
    \bnabla_\perp^2 \delta\! B_\parallel = -\frac{4\upi}{B}\bnabla_\perp\bnabla_\perp : \sum_s M_s \int \mathrm{d}^3\boldsymbol{v}\ \langle \boldsymbol{v}_\perp \boldsymbol{v}_\perp h_s\rangle_r.
\end{equation}

Once all quantities are normalized (see Appendix \ref{normalizations} for normalizations), the non-dimensional gyrokinetic equation is
\begin{equation}
\begin{split}
\pderiv{g_s}{t} + &\left[v_{ts}  v_\parallel \bhat + \gyavgR{\vEM} +
                  \frac{\tau_s}{Z_s}{\boldsymbol v}_d \right] \bcdot \bnabla h_s - v_{ts} \mu \left(\bhat\bcdot\bnabla B\right) \pderiv{h_s}{v_\parallel} \\ 
                  &= - \gyavgR{\vEM} \bcdot \bnabla\big|_E F_{Ms} + C(h_s),
\label{gkg}
\end{split}
\end{equation}
and the non-dimensional field equations are
\begin{gather}
    \sum_s \frac{ Z_s^2 n_s}{\tau_s} \Phi = \sum_s Z_s n_s \int \mathrm{d}^3{\boldsymbol v}\ \gyavgr{h_s}, \label{qneuts} \\
    -\bnabla_\perp^2 A_\parallel =\frac{\beta_\mathrm{ref}}{2}\sum_s Z_s n_s v_{ts}  \int \mathrm{d}^3{\boldsymbol v}\ v_\parallel  \gyavgr{h_s}, \label{ampere}
\end{gather}
and
\begin{gather}
    \bnabla_\perp^2 \delta\! B_\parallel = -\frac{\beta_\mathrm{ref}}{2 {\color{black} B}}\bnabla_\perp\bnabla_\perp : \sum_s n_s \tau_s \int \mathrm{d}^3\boldsymbol{v}\ \langle \boldsymbol{v}_\perp \boldsymbol{v}_\perp h_s\rangle_r, \label{dB}
\end{gather}
where $\beta_{\mathrm{ref}} \equiv 8\upi n_{\mathrm{ref}}T_{\mathrm{ref}}/B_N^2$ is the ratio of the plasma to magnetic pressure for the reference species and  $B_N$ is a reference magnetic field strength.
These equations are solved in a radially-local flux-tube spatial domain that follows a single field line on a single magnetic flux surface.

In {\tt GX} the distribution functions $g_s$ and $h_s$ are both expanded in a Fourier-Hermite-Laguerre series:
\begin{align}
    g_s &= \sum_{\boldsymbol k_\perp}\sum_{\ell} \sum_m \frac{(-1)^{\ell}}{\sqrt{(2\upi)^3m!}} G_{\ell,m}^s\left(\boldsymbol{k}_\perp, z\right) \mathrm{L}_{\ell}(\mu B) \mathrm{He}_m(v_{\parallel})\mathrm{e}^{-v_{\parallel}^2/2 - \mu B} \mathrm{e}^{\mathrm{i} {\boldsymbol k_\perp \bcdot \boldsymbol R}},\\
    h_s &= \sum_{\boldsymbol{k}_\perp}\sum_{\ell} \sum_m \frac{(-1)^{\ell}}{\sqrt{(2\upi)^3m!}} H_{\ell,m}^s\left(\boldsymbol k_\perp, z\right) \mathrm{L}_{\ell}(\mu B) \mathrm{He}_m(v_{\parallel})\mathrm{e}^{-v_{\parallel}^2/2 - \mu B} \mathrm{e}^{\mathrm{i} {\boldsymbol k_\perp \bcdot \boldsymbol R}},
\end{align}
where the spectral coefficients $G_{\ell, m}$ and $H_{\ell, m}$ are related by
\begin{equation}\label{coefficient relation}
    G_{\ell,m}^s({\boldsymbol k_{\perp}},z) =  H_{\ell,m}^s - \frac{Z_s}{\tau_s} \mathcal{J}_\ell^s \Phi \delta_{m 0} + \frac{Z_s v_{ts}}{\tau_s}\mathcal{J}_\ell^s A_\parallel \delta_{m1} - \left(\mathcal{J}_\ell^s + \mathcal{J}_{\ell-1}^s\right){ \frac{\delta\!B_\parallel}{B}} \delta_{m0}.
\end{equation}
A Fourier representation is used for the spatial dimensions perpendicular to the mean magnetic field ($\boldsymbol{k}_\perp$), while Hermite and Laguerre representations are used for the parallel ($v_{\parallel}$) and perpendicular ($\mu B$) velocity domains. A discrete Fourier collocation grid is used for the spatial direction parallel to the field ($z$) so that the parallel derivative $\nabla_{\parallel}$ can be evaluated spectrally. Substituting this expansion into the gyrokinetic equation and taking integrals over velocity space yields differential equations that describe the evolution of the spectral coefficients $G_{\ell,m}$ and $H_{\ell, m}$:
% \begin{equation}\label{glmevolve}
% \begin{split}
%     \pderiv{G_{\ell,m}^s}{t} &+ \mathcal{N}_{\ell,m}^s 
%   + \textcolor{red}{v_{ts} \nabla_\parallel \left( \sqrt{m+1} \, H^s_{\ell,m+1} + \sqrt{m} \, H^s_{\ell,m-1} \right)}
%  \\ &
%   + \textcolor{blue}{v_{ts}  \Big[ - (\ell+1) \, \sqrt{m+1} \, H^s_{\ell,m+1} - \ell \, \sqrt{m+1} \, H^s_{\ell-1,m+1}}
%   \\
%   & \qquad\qquad \textcolor{blue}{+ \ell \, \sqrt{m} \, H^s_{\ell,m-1}
%   + (\ell+1) \, \sqrt{m} \, H^s_{\ell+1,m-1} \Big] \nabla_\parallel \ln B} \\ &
%    + i\frac{\tau_s}{Z_s} \omega_{d}^\kappa \Big[ \sqrt{(m+1)(m+2)} \, H^s_{\ell,m+2} + (2m+1)H^s_{\ell,m} +\sqrt{m (m-1)} \, H^s_{\ell,m-2} \Big]\\ &
%    + i\frac{\tau_s}{Z_s} \omega_{d}^{\bnabla B} \Big[  (\ell+1) \, H^s_{\ell+1,m} + (2\ell+1)H^s_{\ell,m} +\ell \, H^s_{\ell-1,m} \Big]\\ &
%    = \mathcal{D}_{\ell,m}^s + \mathcal{C}_{\ell,m}^{ss}.
% \end{split}
% \end{equation}

\begin{equation}\label{glmevolve}
\begin{split}
    \pderiv{G_{\ell,m}^s}{t} &+ \mathcal{N}_{\ell,m}^s 
  + \textcolor{red}{v_{ts} \nabla_\parallel \left( \sqrt{m+1} \, H^s_{\ell,m+1} + \sqrt{m} \, H^s_{\ell,m-1} \right)}
 \\ &
  + \textcolor{blue}{v_{ts}  \Big[ - (\ell+1) \, \sqrt{m+1} \, H^s_{\ell,m+1} - \ell \, \sqrt{m+1} \, H^s_{\ell-1,m+1}}
  \\
  & \qquad\qquad \textcolor{blue}{+ \ell \, \sqrt{m} \, H^s_{\ell,m-1}
  + (\ell+1) \, \sqrt{m} \, H^s_{\ell+1,m-1} \Big] \nabla_\parallel \ln B} \\ &
   + \mathrm{i}\frac{\tau_s}{Z_s} \omega_{d}^\kappa \Big[ \sqrt{(m+1)(m+2)} \, H^s_{\ell,m+2} + (2m+1)H^s_{\ell,m}
   \\ & \qquad\qquad + \sqrt{m (m-1)} \, H^s_{\ell,m-2} \Big]\\ &
   + \mathrm{i}\frac{\tau_s}{Z_s} \omega_{d}^{\bnabla B} \Big[  (\ell+1) \, H^s_{\ell+1,m} + (2\ell+1)H^s_{\ell,m} +\ell \, H^s_{\ell-1,m} \Big]\\ &
   = \mathcal{D}_{\ell,m}^s + \mathcal{C}_{\ell,m}^{s}.
\end{split}
\end{equation}
In this equation, $\mathcal{N}^s_{\ell,m}$, $\mathcal{D}^s_{\ell,m}$, and $\mathcal{C}^s_{\ell,m}$ represent the nonlinear, drive, and collision terms, respectively. The terms coloured red and blue have a prefactor of $v_{ts}$. For electrons, $v_{ts}$ is approximately 60 times larger than for (deuterium) ions, making these the stiff terms that we will treat implicitly. The red terms correspond physically to parallel streaming along the field line, while the blue terms correspond to the mirror force from the inhomogeneity of the magnetic field strength.

The field equations for $\Phi$, $A_{\parallel}$, and $\delta\!B_\parallel$ are
% \begin{gather}\label{field eqs}
%      \sum_s \frac{ Z_s^2n_s}{\tau_s}\left(1-\sum_{\ell=0}^{\color{black} N_\ell}(\mathcal{J}^s_\ell)^2\right)\Phi - \sum_s Z_s n_s \sum_{\ell=0}^{\color{black} N_\ell}\mathcal{J}^s_\ell(\mathcal{J}^s_\ell+\mathcal{J}^s_{\ell-1}){\color{black} \frac{\delta\! B_\parallel}{B}}  = \sum_s Z_s n_s \sum_{\ell=0}^{\color{black} N_\ell} \mathcal{J}^s_\ell G^s_{\ell,0}, \\
%       \left(k_\perp^2 + \frac{\beta_{\mathrm{ref}}}{2}\sum_s \frac{Z_s^2 n_s}{m_s} \sum_{\ell=0}^{\color{black} N_\ell} (\mathcal{J}^s_\ell)^2\right)A_\parallel = \frac{\beta_{\mathrm{ref}}}{2}\sum_s Z_s n_s v_{ts} \sum_{\ell=0}^{\color{black} N_\ell} \mathcal{J}^s_\ell G^s_{\ell,1},
% \end{gather}
\begin{equation}
\begin{split}
     \sum_s \frac{ Z_s^2n_s}{\tau_s}\left(1-\sum_{\ell=0}^{\color{black} N_\ell}(\mathcal{J}^s_\ell)^2\right)\Phi &- \sum_s Z_s n_s \sum_{\ell=0}^{\color{black} N_\ell}\mathcal{J}^s_\ell(\mathcal{J}^s_\ell+\mathcal{J}^s_{\ell-1}){\color{black} \frac{\delta\! B_\parallel}{B}}  \\
     &= \sum_s Z_s n_s \sum_{\ell=0}^{\color{black} N_\ell} \mathcal{J}^s_\ell G^s_{\ell,0},
\end{split}
\end{equation}
\begin{equation}
     \left(k_\perp^2 + \frac{\beta_{\mathrm{ref}}}{2}\sum_s \frac{Z_s^2 n_s}{m_s} \sum_{\ell=0}^{\color{black} N_\ell} (\mathcal{J}^s_\ell)^2\right)A_\parallel = \frac{\beta_{\mathrm{ref}}}{2}\sum_s Z_s n_s v_{ts} \sum_{\ell=0}^{\color{black} N_\ell} \mathcal{J}^s_\ell G^s_{\ell,1},
\end{equation}
and
\begin{equation}
\begin{split}
     \frac{\beta_{\mathrm{ref}}}{2{\color{black} B^2}}\sum_s Z_s n_s \sum_{\ell=0}^{\color{black} N_\ell}\mathcal{J}^s_\ell(\mathcal{J}^s_\ell+\mathcal{J}^s_{\ell-1})\Phi + &\left(1 + \frac{\beta_{\mathrm{ref}}}{2{\color{black} B^2}}\sum_s n_s \tau_s \sum_{\ell=0}^{\color{black} N_\ell}(\mathcal{J}^s_\ell + \mathcal{J}^s_{\ell-1})^2\right){\color{black} \frac{\delta\! B_\parallel}{B}}\\
     &= -\frac{\beta_{\mathrm{ref}}}{2{\color{black} B^2}}\sum_s n_s \tau_s \sum_{\ell=0}^{\color{black} N_\ell}(\mathcal{J}^s_\ell + \mathcal{J}^s_{\ell-1})G^s_{\ell,0},
\end{split}
\end{equation}
where $\mathcal{J}_{\ell}^s\left(k_{\perp}^2\rho_s^2\right)$ is the Laguerre projection of the Bessel function. In practice, Eq.~\eqref{coefficient relation} is used to replace $H_{\ell,m}$ for $G_{\ell,m}$ in Eq. \eqref{glmevolve}. The resulting equation is then advanced for $G_{\ell,m}$ and the fields are resolved to obtain the updated $H_{\ell,m}$. The details on these derivations can be found in \cite{Mandell2018Laguerre-HermiteGyrokinetics, Mandell2024GX:Design}. Finally, we note that while by default {\tt GX} includes $\delta\!B_{\parallel}$, in this work we neglect parallel magnetic field fluctuations, i.e.  $\delta\!B_{\parallel} = 0$. Ultimately, the implicit algorithms presented can be generalized to include $\delta\!B_{\parallel}$ in a relatively straightforward manner.

\section{Time-Integration Schemes}\label{time integration schemes}
In this work, we use two types of time-integration schemes: a Lie-Trotter splitting scheme \citep{Boyd2001ChebyshevMethods} similar to the one used in {\tt stella} \citep{Barnes2019Stella:Configurations}, as well as additive Runge-Kutta methods \citep{Cooper1980AdditiveEquations}.

\subsection{Lie-Trotter}
We begin with the Lie-Trotter scheme. For simplicity, we write Eq. \eqref{glmevolve} as
\begin{equation}
    \label{glmlietrotter}
    \frac{\partial\boldsymbol{G}}{\partial t} = A[\boldsymbol{G}] + \mathsfbi{B}\boldsymbol{G} + \mathsfbi{C}\boldsymbol{G},
\end{equation}
where $\boldsymbol{G}$ is a vector containing all of the spectral coefficients of all species and the linear operators $\mathsfbi{B}$ and $\mathsfbi{C}$ represent the stiff streaming and mirror operators
\begin{equation}
     \left(\mathsfbi{B}\boldsymbol{G}\right)_{\ell,m} = v_{ts} \nabla_\parallel \left( \sqrt{m+1} \, H^s_{\ell,m+1} + \sqrt{m} \, H^s_{\ell,m-1} \right),
\end{equation}
and
\begin{equation}
\begin{split}
  \left(\mathsfbi{C}\boldsymbol{G}\right)_{\ell,m} = v_{ts} \Big[ &- (\ell+1) \, \sqrt{m+1} \, H^s_{\ell,m+1} - \ell \, \sqrt{m+1} \, H^s_{\ell-1,m+1} \\
  &+ \ell \, \sqrt{m} \, H^s_{\ell,m-1} + (\ell+1) \, \sqrt{m} \, H^s_{\ell+1,m-1} \Big] \nabla_\parallel \ln B,
\end{split}
\end{equation}
respectively.
The remaining terms, including the nonlinear term, are contained in the operator $A$. In a Lie-Trotter splitting, we advance the spectral coefficients $\boldsymbol{G}^n$ at time $t^n$ to $\boldsymbol{G}^{n+1}$ at $t^{n+1}$ by evaluating each operator sequentially:
\begin{gather}
     \boldsymbol{G}^{\overline{n}} = \boldsymbol{G}^{n} + \Delta t A[\boldsymbol{G}^n], \\
     \boldsymbol{G}^{\overline{\overline{n}}} = \boldsymbol{G}^{\overline{n}} + \Delta t \mathsfbi{B}\boldsymbol{G}^{\overline{\overline{n}}}, \\
     \boldsymbol{G}^{n+1} = \boldsymbol{G}^{\overline{\overline{n}}} + \Delta t \mathsfbi{C}\boldsymbol{G}^{n+1}.
\end{gather}

For the explicit operator $A$, we solve for the intermediate $\boldsymbol{G}^{\overline{n}}$ using $\boldsymbol{G}^n$ at the current timestep. However, the equations containing $\mathsfbi{B}$ and $\mathsfbi{C}$ use the second, intermediate, and future stage, and this results in two matrix equations that must be solved. Furthermore, this method does not require storing extra copies of $\boldsymbol{G}$, which makes it very memory-efficient.

While in the above equations, the explicit operator is advanced using a Forward Euler scheme, we in fact follow \citet{Barnes2019Stella:Configurations} and use the standard third-order Strong-Stability Preserving Runge-Kutta (SSPRK3) method from \citet{Gottlieb2001StrongMethods}. The implicit advances are performed using the implicit method from \cite{Giraldo2013Implicit-ExplicitNUMA}, which is second-order accurate. We find that the use of a second-order implicit scheme is important for accurately capturing transitions between different types of linear instabilities in toroidal geometry even though the entire scheme taken as a whole is first order. Strang-splitting is commonly used to obtain second-order accuracy at a higher computational cost. Nevertheless, it has been observed that the asymptotic convergence rate may only be realized for sufficiently small timesteps, and that the associated error constant can be large \citep{Ropp2004StudiesEquations}. This motivates the use of higher-order IMEX methods.

\subsection{Additive Runge-Kutta}
In the additive Runge-Kutta (ARK) methods, we solve for the solution at the next timestep $\boldsymbol{G}^{n+1}$ using a series of stages
\begin{align}
    \boldsymbol{y}^{(i)} = \boldsymbol{G}^n + \Delta t \sum_j a_{ij} A\left[\boldsymbol{y}^{(j)}\right] + \Delta t \sum_j \tilde{a}_{ij} \mathsfbi{B}^*\boldsymbol{y}^{(j)},
\end{align}
and
\begin{align}
    \boldsymbol{G}^{n+1} = \boldsymbol{G}^n + \Delta t \sum_j b_{j} A\left[\boldsymbol{y}^{(j)}\right] + \Delta t \sum_j \tilde{b}_{j} \mathsfbi{B}^*\boldsymbol{y}^{(j)}.
\end{align}
The operator $A$ is the explicit operator as for the Lie-Trotter scheme, but now $\mathsfbi{B}^*$ contains both the streaming and mirror terms
% \begin{equation}
% \begin{split}
%     \left(B^* \boldsymbol{G}\right)_{\ell,m} = v_{ts}\Big[ \nabla_\parallel \left( \sqrt{m+1} \, H^s_{\ell,m+1} + \sqrt{m} \, H^s_{\ell,m-1} \right) \\
%     \Big(- (\ell+1) \, \sqrt{m+1} \, H^s_{\ell,m+1} - \ell \, \sqrt{m+1} \, H^s_{\ell-1,m+1} \\
%     + \ell \, \sqrt{m} \, H^s_{\ell,m-1}
%   + (\ell+1) \, \sqrt{m} \, H^s_{\ell+1,m-1}\Big) \nabla_\parallel \ln B \Big].
% \end{split}
% \end{equation}
\begin{equation}
\begin{split}
  \left(\mathsfbi{B}^* \boldsymbol{G}\right)_{\ell,m} = v_{ts}\Big[ &\nabla_\parallel \left( \sqrt{m+1} \, H^s_{\ell,m+1} + \sqrt{m} \, H^s_{\ell,m-1} \right) \\
  &+ \Big(- (\ell+1) \, \sqrt{m+1} \, H^s_{\ell,m+1} - \ell \, \sqrt{m+1} \, H^s_{\ell-1,m+1} \\
  &\qquad + \ell \, \sqrt{m} \, H^s_{\ell,m-1} + (\ell+1) \, \sqrt{m} \, H^s_{\ell+1,m-1}\Big) \nabla_\parallel \ln B \Big].
\end{split}
\end{equation}

There are many different choices for coefficients $a_{ij}$, $\Tilde{a}_{ij}$, $b_i$ and $\Tilde{b}_i$ in the published literature. For this work, we use the coefficients from the `ARK3(2)4L[2]SA–ESDIRK' method of \citet{Kennedy2003AdditiveEquations}. This method contains 4 stages and is overall 3rd order accurate. Although the scheme has a higher order of accuracy than the Lie-Trotter method, it requires storing the intermediate solutions $\boldsymbol{y}^{(i)}$ and performing three linear solves per step.

A key challenge for implicit gyrokinetic simulations is that the implicit operators act on both the plasma distribution function and the electromagnetic fields (which are both contained in $H_{\ell,m}$), which are themselves obtained from integral operators of the distribution function. This coupling makes the inversion significantly more difficult. In the next section, we describe several algorithms designed to address this difficulty.

\section{Solving the Implicit Time-Advance Equations}\label{advance implicit}
In this section, we describe our methods to invert the parallel streaming and mirror force operators for our implicit methods. In  \S\ref{k's method}, we discuss Kotschenreuther's method, which is used to invert the streaming operator in {\tt GS2} and {\tt stella}. Due to the large memory requirements of this method, we introduce new methods in \S\ref{invert stream} and \S\ref{invert stream with apar} that approximately invert the streaming operator. Then in \S\ref{invert mirror}, we present our method for inverting the mirror term. Sections \ref{invert hypercollisions} and \ref{remaining terms} discuss the hypercollision and magnetic/curvature drift terms and their stiffness. Due to approximations made in inverting the streaming and mirror operators, in \S\ref{stationary iteration} and \S\ref{operator split preconditioning} we discuss simple iterative methods to improve the accuracy of our methods.
\subsection{Kotschenreuther's Method}\label{k's method}
We start by briefly describing Kotschenreuther's method \citep{Kotschenreuther1995ComparisonInstabilities} for inverting the linear operators in the gyrokinetic equation.
The Kotschenreuther method is used to invert all the linear terms in {\tt GS2}, as well as the streaming terms for ${\tt stella}$.

Projecting the parallel streaming terms onto the Hermite-Laguerre basis yields the equation
\begin{equation}
\pderiv{G_{\ell,m}^s}{t} + v_{ts} \nabla_\parallel \left( \sqrt{m+1} \, G_{\ell,m+1}^s + \sqrt{m} \, G_{\ell,m-1}^s + \frac{Z_s}{\tau_s}\mathcal{J}_\ell^s \Phi \delta_{m-1,0}\right) = 0,
\end{equation}
where for simplicity we have ignored the drive and collision terms. We have also assumed only electrostatic perturbations.
Discretizing in time using a Backwards Euler scheme, we must solve an equation of the form
\begin{equation}\label{Kot stream}
\begin{split}
    G_{\ell,m}^{s,n+1} &+ \Delta t\, v_{ts} \nabla_{\parallel}\left(\sqrt{m+1}\,G_{\ell,m+1}^{s,n+1} + \sqrt{m}\,G_{\ell,m-1}^{s,n+1}\right) \\
    &+ \Delta t\,\frac{Z_s}{\tau_s}v_{ts}\delta_{m1}\nabla_{\parallel}\left(\mathcal{J}_{\ell}^s\Phi^{n+1}\right) = G_{\ell,m}^{s,n},
\end{split}
\end{equation}
where the $n$ and $n+1$ indices indicate the current and future timesteps, respectively.
In {\tt GX}, the above equation is then Fourier transformed in the $z$ domain, resulting in the equation
\begin{equation}\label{Kot stream fft}
\begin{split}
    \hat{G}_{\ell,m}^{s,n+1} &+ \Delta t v_{ts} \mathrm{i} k_{\parallel}\left(\sqrt{m+1}\hat{G}_{\ell,m+1}^{s,n+1} + \sqrt{m}\hat{G}_{\ell,m-1}^{s,n+1}\right) \\
    &+ \Delta t\frac{Z_s}{\tau_s}v_{ts}\delta_{m1}\mathrm{i}k_{\parallel}\mathcal{F}\left[\mathcal{J}_{\ell}^s\Phi^{n+1}\right] = \hat{G}_{\ell,m}^{s,n},
\end{split}
\end{equation}
where the hats and $\mathcal{F}[\ ]$ denote Fourier transforms.

We separate the solution $\hat{G}_{\ell,m}^s$ into a homogeneous and inhomogeneous part $\hat{G}_{\ell,m}^s = \hat{G}_{\ell,m}^{h,s} + \hat{G}_{\ell,m}^{i,s}$ satisfying
\begin{equation}
\label{homogenous}
\begin{split}
    \hat{G}_{\ell,m}^{h,s,n+1} &+ \Delta t v_{ts} \mathrm{i} k_{\parallel}\left(\sqrt{m+1}\hat{G}_{\ell,m+1}^{h,s,n+1} + \sqrt{m}\hat{G}_{\ell,m-1}^{h,s,n+1}\right) \\ &+  \Delta t\frac{Z_s}{\tau_s}v_{ts}\delta_{m1}\mathrm{i} k_{\parallel}\mathcal{F}\left[\mathcal{J}_{\ell}^s\Phi^{n+1}\right] = 0,
\end{split}
\end{equation}
and
\begin{equation}\label{inhomogenous}
    \hat{G}_{\ell,m}^{i,s,n+1} + \Delta t v_{ts} \mathrm{i}k_{\parallel}\left(\sqrt{m+1}\hat{G}_{\ell,m+1}^{i,s,n+1} + \sqrt{m}\hat{G}_{\ell,m-1}^{i,s,n+1}\right) = \hat{G}_{\ell,m}^{s,n}.
\end{equation}
We can straightforwardly solve Eq. \eqref{inhomogenous} for $\hat{G}_{\ell,m}^{i,s}$ using a tri-diagonal solver. Specifically, Eq. \eqref{inhomogenous} is tridiagonal with respect to the $m$ indices, and the solver concurrently solves the equation across the remaining parameters. To solve for the homogeneous part, we need to first find an expression for $\Phi$. Note that $\Phi$ can be expressed as
% \begin{equation}
%     D\Phi = \sum_s Z_s n_s \sum_\ell \mathcal{J}_\ell \left(G_{\ell,0}^{i,s} + \sum_p^{N_z}\frac{\delta G_{\ell,0}^{h,s}(z_i)}{\delta \Phi_p}\Phi\right)
% \end{equation}
\begin{equation}
    \left[\sum_s \frac{Z_s^2n_s}{\tau_s}\left(1-\sum_{\ell=0}^{\color{black} N_\ell}(\mathcal{J}^s_\ell)^2\right)\right]\Phi = \Phi^i + \left(\sum_s Z_s n_s \sum_\ell \mathcal{J}_\ell \mathsfbi{M}_{\ell}^s\right)\Phi,
\end{equation}
where $\Phi^i$ is the contribution to $\Phi$ from the inhomogeneous solution, and $\mathsfbi{M}_{\ell}^s = \delta G_{\ell,0}^{h,s}(z_i)/\delta \Phi_j$ is called the `response matrix'. $\mathsfbi{M}_{\ell,i,j}^s$ is the response at $z_i$, in $G_{\ell,0}^{h,s}(z_i)$ due to a unit perturbation in $\Phi$ at $z_j$. The response matrix can be calculated for each $z$, $k_x$, and $k_y$ using another series of tridiagonal matrix solves. For each solve, we set $\Phi=\delta_{ij}$, and multiply it by $\mathcal{J}_\ell^s$ before Fourier transforming.
From there, we can solve for $\Phi$ by solving
\begin{equation}
    \left(\mathsfbi{I} - D^{-1}\sum_s Z_s n_s \sum_{\ell} \mathcal{J}_{\ell}^s \mathsfbi{M}\right)\Phi = \Phi^i,
\end{equation}
where $D = \sum_s Z_s^2n_s/\tau_s\left(1-\sum_{\ell=0}^{\color{black} N_\ell}(\mathcal{J}^s_\ell)^2\right)$. We can then substitute $\Phi$ into the original equation and solve for $G_{\ell,m}^s$.

Although exact, the major issue with using Kotschenreuther's method for {\tt GX} is the memory requirement. For $N_z$ points along the field-line and $N_x$ and $N_y$ number of $k_x$ and $k_y$ wavenumbers, there are $N_x \, N_y$ matrices of dimension $N_z$, each of which needs to be inverted to solve for $\Phi$. However, when including boundary conditions for the flux-tube, the dimension of each matrix is now $N_z \, N_{\mathrm{links}}$, where $N_{\mathrm{links}}$ is the number of flux tubes at different $k_x$ that are linked together. For a moderate resolution nonlinear simulation, $N_z \approx 32$, $N_x \approx N_y \approx 256$, and $N_{\mathrm{links}} \approx 10$. As ${\tt GX}$ uses single-precision complex values for all of its data, the total memory required could then reach approximately 100 GB, which would be well beyond the amount of memory most GPUs today contain, especially if the same GPU must also contain all of the moments and fields. These matrices will also be dense, and so cannot be efficiently handled with sparse matrix solvers. Finally, the need to compute and store the response matrices makes this method less amenable to matrix-free iterative methods. This motivates the need for alternative approaches such as the ones described in this paper.

\subsection{Parallel Streaming}\label{invert stream}
As above, we would like to solve an equation of the form
\begin{equation}\label{streaming fourier}
\begin{split}
\hat{G}_{\ell,m}^{s,n+1} &+ \Delta t v_{ts} \mathrm{i} k_{\parallel}\left(\sqrt{m+1}\hat{G}_{\ell,m+1}^{s,n+1} + \sqrt{m}\hat{G}_{\ell,m-1}^{s,n+1}\right) \\
&+ \Delta t\frac{Z_s}{\tau_s}v_{ts}\delta_{m1}\mathrm{i} k_{\parallel}\mathcal{F}\left[\mathcal{J}_{\ell}^s\Phi^{n+1}\right] = \hat{G}_{\ell,m}^{s,n}.
\end{split}
\end{equation}
Without $\Phi$, the equation is tridiagonal in the Hermite moments $m$. However, the term $\mathcal{F}\left[\mathcal{J}_\ell^s \Phi\right]$ involves convolving $\Phi$, which makes the equation difficult to solve. Instead of directly inverting, we will use an iterative scheme to solve this equation.

Assuming only a single ion species and electrons, the equation for $\Phi$ is
\begin{equation}
     \Phi = \frac{Z_i n_i \sum_\ell \mathcal{J}^i_\ell G_{\ell 0}^{i}}{\sum_s \frac{Z_s^2 n_s}{\tau_s}\left(1-\sum_\ell \left(\mathcal{J}^s_\ell\right)^2\right)} + \frac{Z_e n_e \sum_\ell \mathcal{J}^e_\ell G_{\ell 0}^{e}}{\sum_s \frac{Z_s^2 n_s}{\tau_s}\left(1-\sum_\ell \left(\mathcal{J}^s_\ell\right)^2\right)}.
\end{equation}
To simplify this equation, we first take a drift-kinetic ($k_\perp \rho_i \ll 1$) limit. This simply results in neglecting all finite-Larmor-radius (FLR) terms in the equation for $\Phi$ by setting $\mathcal{J}^s_0 = 1$ and $\mathcal{J}^s_{\ell > 0} = 0$ except for in the polarization term  $\sum_s Z_s^2n_s/\tau_s\left(1-\sum_{\ell=0}^{\color{black} N_\ell}(\mathcal{J}^s_\ell)^2\right)$. The $\mathcal{J}^s_\ell$ multiplying $\Phi$ in Eq. \eqref{streaming fourier} is also set to 1. The equation for $\Phi$ is then
\begin{equation}
     \Phi \approx \frac{Z_i n_i G_{0,0}^{i}}{\sum_s \frac{Z_s^2 n_s}{\tau_s}\left(1-\sum_\ell \left(\mathcal{J}^s_\ell\right)^2\right)} + \frac{Z_e n_e G_{0, 0}^{e}}{\sum_s \frac{Z_s^2 n_s}{\tau_s}\left(1-\sum_\ell \left(\mathcal{J}_\ell^s\right)^2\right)}.
\end{equation}
However, the denominator is still a function of the parallel coordinate, and would still result in a convolution after Fourier-transforming. To avoid this, we approximate the denominator by its minimum over $z$:
\begin{equation}
     \Phi = \frac{Z_i n_i G_{0,0}^{i}}{D_m} + \frac{Z_e n_e G_{0, 0}^{e}}{D_m},
\end{equation}
where $D_m$ is
\begin{equation}
    D_m = \min_{z} \left[\sum_s \frac{Z_s^2 n_s}{\tau_s}\left(1-\sum_\ell \left(\mathcal{J}^s_\ell\right)^2\right)\right].
\end{equation}
As described in Appendix \ref{app a}, we make this choice to capture the highest frequency wave in the system. An iterative method described in \S\ref{stationary iteration} will then help accurately capture the remaining modes.

Substituting this approximation for $\Phi$ into Eq. \eqref{streaming fourier} yields
\begin{equation}
\label{streaming fourier approx}
\begin{split}
\hat{G}_{\ell,m}^{s,n+1} &+ \Delta tv_{ts} \mathrm{i} k_{\parallel} \left( \sqrt{m+1} \, \hat{G}_{\ell,m+1}^{s,n+1} + \sqrt{m} \, \hat{G}_{\ell,m-1}^{s,n+1}\right) \\&+\Delta tv_{ts} \mathrm{i} k_{\parallel} \frac{Z_s}{\tau_s} \left(\frac{Z_i n_i \hat{G}_{0,0}^{i,n+1}}{D_m} + \frac{Z_e n_e \hat{G}_{0, 0}^{e,n+1}}{D_m} \right)\delta_{\ell 0}\delta_{m 1}
= \hat{G}_{\ell,m}^{s,n}.
\end{split}
\end{equation}
This equation is now nearly tridiagonal, with simple corrections from $\Phi$. The resulting matrix system can be written as
\begin{equation}
    \left(\mathsfbi{B}_T + \boldsymbol{u} \otimes \boldsymbol{v}\right)\boldsymbol{G} = \boldsymbol{r},
\end{equation}
where the matrix $\mathsfbi{B}_T$ contains the tridiagonal bands, and the outer product of $\boldsymbol{u}$ and $\boldsymbol{v}$ contains the approximation to $\Phi$. The expressions for $\boldsymbol{u}$ and $\boldsymbol{v}$ are
\begin{gather}
    \boldsymbol{u} =\mathrm{i}k_{\parallel}\Delta t  \left[0 \ \frac{Z_i v_{ti}}{\tau_i} \delta_{\ell 0}\delta_{m 1}\ \boldsymbol{0} \ \frac{Z_e v_{te}}{\tau_e}\delta_{\ell 0}\delta_{m 1} \ \boldsymbol{0}\right],\label{sm phi u}
\end{gather}
and 
\begin{gather}
    \boldsymbol{v} = \left[\frac{Z_in_i}{D_m}\delta_{\ell 0}\delta_{m 0} \ \boldsymbol{0} \ \frac{Z_en_e}{D_m}\delta_{\ell 0}\delta_{m 0} \ \boldsymbol{0}\right]\label{sm phi v},
\end{gather}
where the bolded $\boldsymbol{0}$ indicates consecutive zeros.

\begin{figure}
    \centering
    \includegraphics[width=0.9\linewidth]{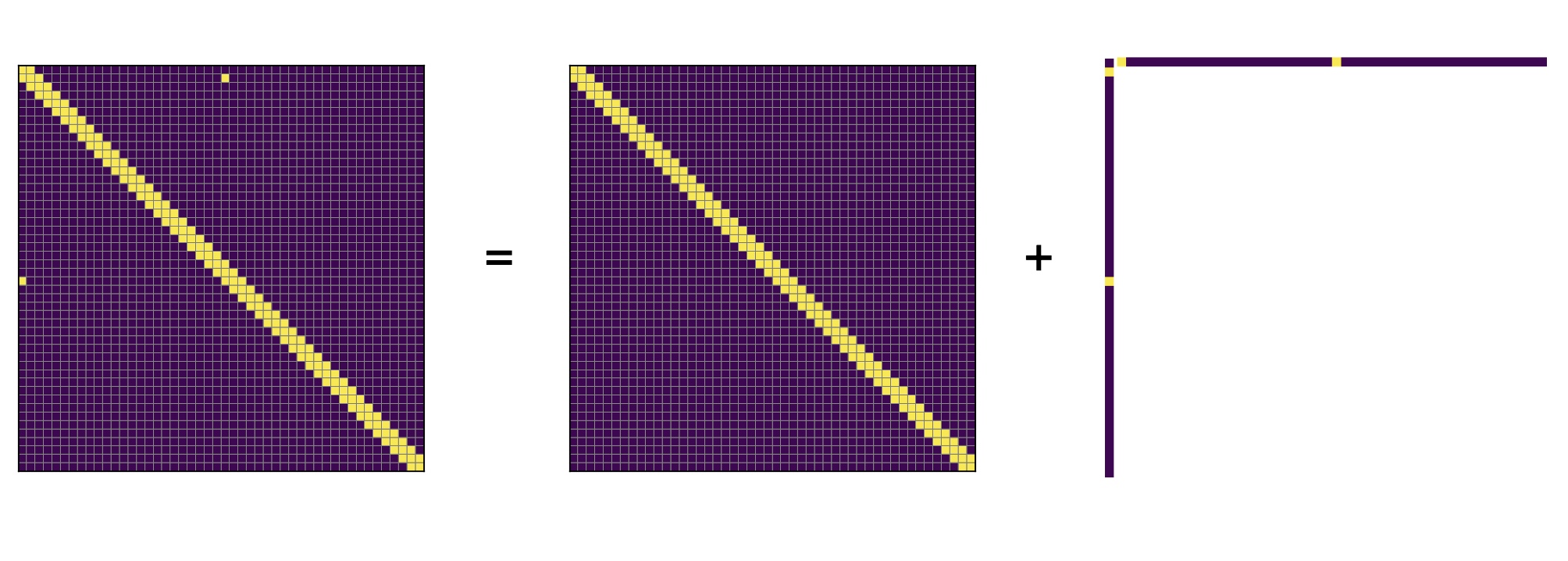}
    \caption{A visualization of the banded structure of the streaming matrix, with only electrostatic perturbations.}
    \label{streaming es banded structure}
\end{figure}

Since $\Phi$ is a low-rank correction, we can solve the above system using the Sherman-Morrison formula, as detailed in Appendix \ref{app sherman morrison}. A visualization of this banded structure is shown in figure \ref{streaming es banded structure}.

It should be noted that we did not have to take a long-wavelength approximation for $\Phi$, and it would still be a low-rank correction. However, by taking a long-wavelength approximation, the intermediate solutions needed for the Sherman-Morrison formula are only of dimension $n_s \, L \, M \, N_z$, where $n_s$ is the number of species, $L$ is the number of Laguerre moments, and $M$ is the number of Hermite moments. If we did not take a long-wavelength approximation, then the size would be $n_s \, L \, M \, N_z \, N_y \, N_x$. This memory saving is important given the limited memory on GPUs. Overall, this method involves only a small number of tri-diagonal solves, which can be computed in linear time without extra storage, as well as a series of cheap vector operations where the vectors are also low in memory. Furthermore, as we explain in \S\ref{stationary iteration}, this approximation is only used as a preconditioner for an iterative method. For the simulations in this work, we only focus on ion-scale turbulence in which $k_{\perp}\rho_i \lesssim 1$, and the iterative method converges rapidly. However, it is possible that shorter wavelength instabilities would require a more accurate preconditioner, and this will be addressed in future work.

When simulating multiple $k_x$ wavenumbers in the presence of magnetic shear, $\hat{s} \ne 0$, enforcing periodicity in $x$ results in the well-known `twist-and-shift' boundary conditions \citep{Beer1995GyrofluidTokamaks, Martin2018TheDevices}. Modes with differing $k_x$ (within the grid) are coupled at the end of the  $z$ domain via the formula
\begin{equation}
    {\color{black} H_{\ell,m}(k_x, k_y, z) = H_{\ell,m}(k_x+\delta k, k_y, z+2\upi)},
\end{equation}
where $\delta k = 2\upi k_y \hat{s}$. One can interpret this formula as simulating an extended flux-tube that contains all the coupled $k_x$'s. The above algorithm remains mostly unchanged in this new flux-tube, except that the Fourier transforms in $z$ are taken at fixed $k_y$ along this extended flux tube. Similarly, we take $D_m$ to be the minimum along the extended flux-tube:
\begin{equation}
    D_m(k_y) = \min_{\forall k_{x}, z} D(z).
\end{equation}

\subsection{Streaming with $A_{\parallel}$}\label{invert stream with apar}

When including $A_{\parallel}$, the parallel streaming terms become
\begin{equation}
\begin{split}
\pderiv{G_{\ell,m}^s}{t} +& v_{ts} \nabla_\parallel \Big[ \sqrt{m+1} \, G_{\ell,m+1}^s + \sqrt{m} \, G_{\ell,m-1}^s + \frac{Z_s}{\tau_s}\mathcal{J}_\ell^s \Phi \delta_{m 1} \\
&- \frac{Z_s v_{ts}}{\tau_s}\mathcal{J}_{\ell}^s A_{\parallel}\left(\delta_{m 0} + \sqrt{2}\delta_{m 2}\right)\Big] = r.
\end{split}
\end{equation}
Like with $\Phi$, we take a drift-kinetic limit. In this limit $A_{\parallel}$ is simplified to
\begin{equation}
    A_{\parallel} = \frac{\beta_{\mathrm{ref}}}{2F}\left(Z_i n_i v_{ti} G_{0,1}^i + Z_e n_e v_{te} G_{0,1}^e\right),
\end{equation}
and $F$ is
\begin{equation}
    F = k_\perp^2 + \frac{\beta_{\mathrm{ref}}}{2}\sum_s \frac{Z_s^2 n_s}{m_s} \sum_{\ell=0}^{\color{black} N_\ell} (\mathcal{J}^s_\ell)^2.
\end{equation}
As described in Appendix \ref{app a}, we maximize rather than minimize $F$:
\begin{equation}
    F_M = \max_z\left[k_\perp^2 + \frac{\beta_{\mathrm{ref}}}{2}\sum_s \frac{Z_s^2 n_s}{m_s} \sum_{\ell=0}^{\color{black} N_\ell} (\mathcal{J}^s_\ell)^2\right].
\end{equation}
Our final approximation for $A_{\parallel}$ is thus
\begin{equation}\label{Apar dk}
    A_{\parallel} = \frac{\beta_{\mathrm{ref}}}{2F_M}\left(Z_i n_i v_{ti} G_{0,1}^i + Z_e n_e v_{te} G_{0,1}^e\right).
\end{equation}
The final equation that we solve is
\begin{equation}
    \left(\mathsfbi{B}_T + \boldsymbol{u}_1 \otimes \boldsymbol{v}_1 +  \boldsymbol{u}_2 \otimes \boldsymbol{v}_2\right)\boldsymbol{G} = \boldsymbol{r},
\end{equation}
where $\boldsymbol{u}_1$ and $\boldsymbol{v}_1$ are the same as in Eqs. \eqref{sm phi u} and \eqref{sm phi v}, and $\boldsymbol{u}_2$ and $\boldsymbol{v}_2$ are given by
% \begin{equation}
%      \boldsymbol{u}_2 = \left[\frac{Z_iv_{ti}}{\tau_i}\delta_{\ell,0}\delta_{m,0} \ \ 0 \ \ \sqrt{2}\frac{Z_iv_{ti}}{\tau_i}\delta_{\ell,0}\delta_{m,2} \ \ \boldsymbol{0} \ \frac{Z_ev_{te}}{\tau_e}\delta_{\ell,0}\delta_{m,0} \ \ 0 \ \ \sqrt{2}\frac{Z_ev_{te}}{\tau_e}\delta_{\ell,0}\delta_{m,2} \ \boldsymbol{0}\right], \label{sm apar u}
% \end{equation}
\begin{equation}\label{sm apar u}
\begin{split}
    \boldsymbol{u}_2 = -\mathrm{i}k_{\parallel} \Delta t\Big[ &\frac{Z_iv_{ti}^2}{\tau_i}\delta_{\ell 0}\delta_{m 0} \ \ 0 \ \ \sqrt{2}\,\frac{Z_iv_{ti}^2}{\tau_i}\delta_{\ell 0}\delta_{m 2} \ \ \boldsymbol{0} \\
    &\frac{Z_ev_{te}^2}{\tau_e}\delta_{\ell 0}\delta_{m 0} \ \ 0 \ \ \sqrt{2}\,\frac{Z_ev_{te}^2}{\tau_e}\delta_{\ell 0}\delta_{m 2} \ \ \boldsymbol{0}\Big],
\end{split}
\end{equation}
and
\begin{equation}
    \boldsymbol{v}_2 = \left[0 \ \ \frac{\beta_{\mathrm{ref}}}{2F_M}Z_in_iv_{ti}\delta_{\ell 0}\delta_{m 1} \ \ \boldsymbol{0} \ \ \frac{\beta_{\mathrm{ref}}}{2F_M}Z_en_ev_{te}\delta_{\ell 0}\delta_{m 1} \  \boldsymbol{0}\right]\label{sm apar v}.
\end{equation}

\subsection{Mirror Force}\label{invert mirror}
\begin{figure}
     \centering
     \begin{subfigure}[c]{0.25\textwidth}
         \centering         \includegraphics[width=\textwidth]{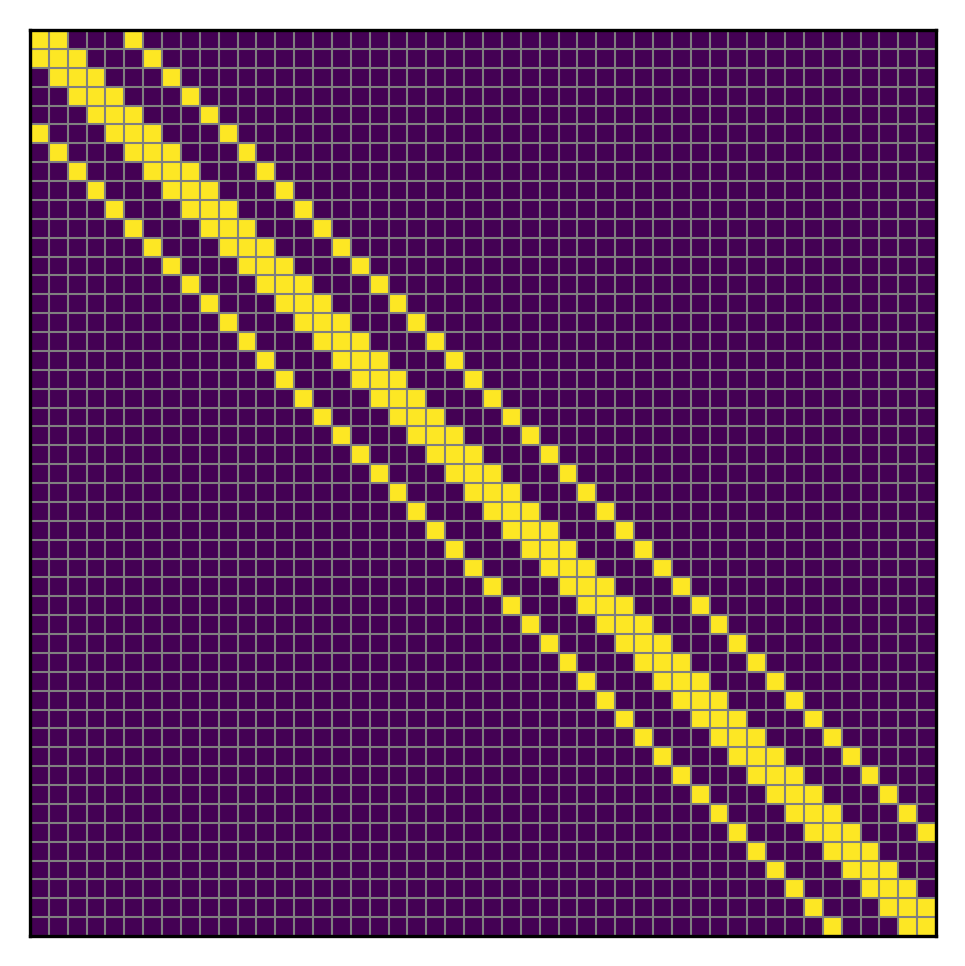}
         \caption{}
         \label{mirror banded structure}
     \end{subfigure}
     \hfill
     \begin{subfigure}[c]{0.25\textwidth}
         \centering         \includegraphics[width=\textwidth]{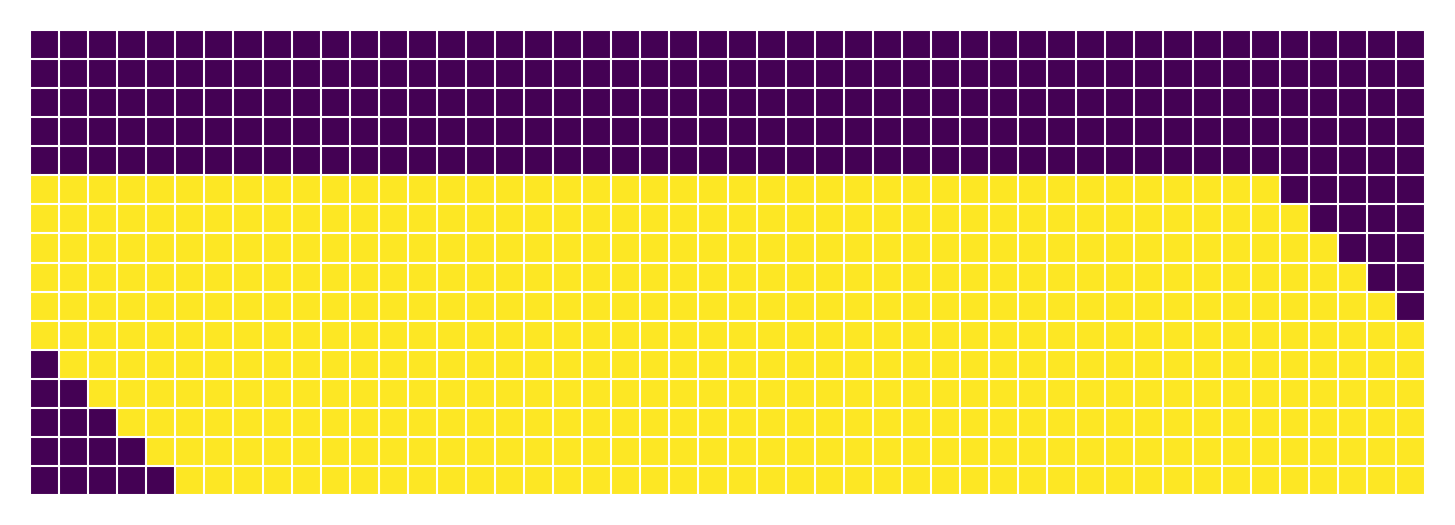}
         \caption{}
         \label{mirror banded structure gbsv}
     \end{subfigure}
     \hfill
     \begin{subfigure}[c]{0.25\textwidth}
         \centering
         \includegraphics[width=\textwidth]{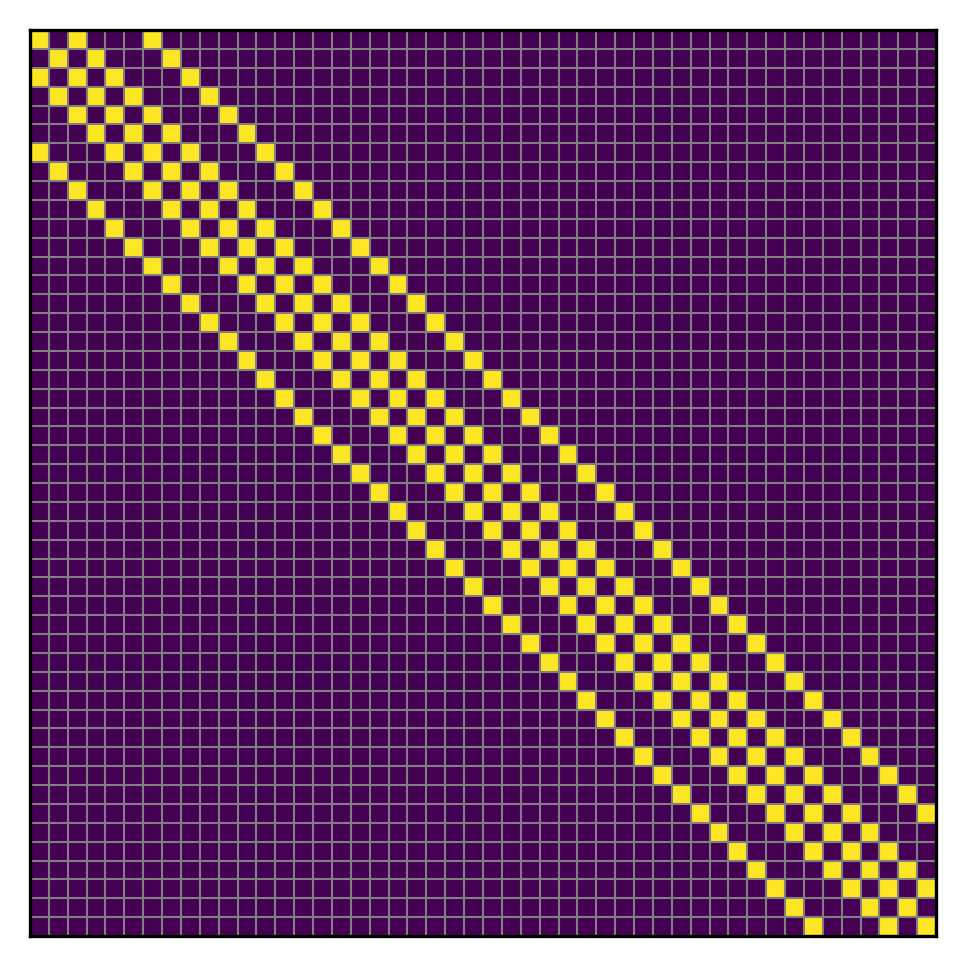}
         \caption{}
         \label{drifts banded structure}
     \end{subfigure}
     \hfill
        \caption{The visualizations of the banded structure of the mirror (a) and magnetic drift (c) terms. The middle figure (b) shows the structure of the mirror matrix that is used in the MAGMA banded matrix routines.}
        \label{mirror and drifts banded structure}
\end{figure}

For the mirror terms, since there is no $\nabla_{\parallel}$ operator, we do not perform any parallel Fourier transforms. As a result, we do not make approximations for the fields and instead directly solve  the equations that arise from our implicit scheme. Furthermore, we will only treat the electron mirror term implicitly. After discretizing in time, the equation involving the electron mirror term is of the form
% \begin{equation}
%     G_{\ell,m}^{s,n+1} + \Delta t v_{ts}\Big[ - (\ell+1) \, \sqrt{m+1} \, H^s_{\ell,m+1} - \ell \, \sqrt{m+1} \, H^s_{\ell-1,m+1} + \ell \, \sqrt{m} \, H^s_{\ell,m-1}\Big] \nabla_\parallel \ln B
%   % + (\ell+1) \, \sqrt{m} \, H^s_{\ell+1,m-1} \Big] 
%   %   \textcolor{blue}{v_{ts}  \Big[ - (\ell+1) \, \sqrt{m+1} \, H^s_{\ell,m+1} - \ell \, \sqrt{m+1} \, H^s_{\ell-1,m+1}} \notag
%   % \\
%   % & \qquad\qquad \textcolor{blue}{+ \ell \, \sqrt{m} \, H^s_{\ell,m-1}
%   % + (\ell+1) \, \sqrt{m} \, H^s_{\ell+1,m-1} \Big] \nabla_\parallel \ln B}
%   % \notag
% \end{equation}
\begin{equation}
\begin{split}
    G_{\ell,m}^{e,n+1} 
  &+ \Delta t{v_{te}  \Big[ - (\ell+1) \, \sqrt{m+1} \, H^{e,n+1}_{\ell,m+1} - \ell \, \sqrt{m+1} \, H^{e,n+1}_{\ell-1,m+1}}
  \\
  & \qquad\qquad {+ \ell \, \sqrt{m} \, H^{e,n+1}_{\ell,m-1}
  + (\ell+1) \, \sqrt{m} \, H^{e,n+1}_{\ell+1,m-1} \Big] \nabla_\parallel \ln B}
 = G_{\ell,m}^{e,n} .
\end{split}
\end{equation}
The matrix we invert is again sparse and banded, with low rank corrections from the fields that are handled using the Sherman-Morrison formula. The banded pattern is shown in figure~\ref{mirror banded structure}. There are 5 bands in the matrix: a tridiagonal structure with two additional bands at an offset of $M-1$ from the main diagonal.

To take advantage of this sparsity pattern, we utilize direct solvers for banded matrices recently developed in the MAGMA linear algebra package \citep{Abdelfattah2023GPU-basedStructure}. In this solver, the original input matrix of size $LM \times LM$ is stored in banded format into a matrix of size $(2K_\mathrm{L} + K_\mathrm{U} + 1) \times LM$, where $K_\mathrm{L}$ and $K_\mathrm{U}$ are the lower and upper bandwidth, respectively. Row $K_\mathrm{L} + K_\mathrm{U} + 1$ of this new matrix contains the main diagonal, while the rows above and below contain the upper and lower diagonals. This structure is shown in figure \ref{mirror banded structure gbsv}. The full mirror matrix contains $4 \, L^2 \, M^2$ entries for 2 species, whereas this condensed matrix only contains $(3 \, M - 2)(2\, L \, M) \sim 6 \, L \, M^2$ entries, and so results in about a factor of $L$ reduction in memory.

To analyse how $\Phi$ enters into the mirror terms, we rewrite each $H_{\ell,m}^e$ in terms of $G_{\ell,m}^e$
\begin{align}
    H_{\ell,m+1}^e &= G_{\ell,m+1}^e - \frac{1}{\tau_e}\mathcal{J}_{\ell}^e\Phi \delta_{m+1,0}, \\
    H_{\ell-1,m+1}^e &= G_{\ell-1,m+1}^e - \frac{1}{\tau_e}\mathcal{J}_{\ell-1}^e\Phi \delta_{m+1,0}, \\
    H_{\ell,m-1}^e &= G_{\ell,m-1}^e - \frac{1}{\tau_e}\mathcal{J}_{\ell}^e\Phi \delta_{m-1,0},\\
    H_{\ell+1,m-1}^e &= G_{\ell+1,m-1}^e - \frac{1}{\tau_e}\mathcal{J}_{\ell+1}^e\Phi \delta_{m-1,0}.
\end{align}
The only contributions are from $\delta_{m-1,0}$ or $\delta_{m,1}$. However, from Eq. \eqref{glmevolve} these terms have prefactors that vanish when taking a drift-kinetic limit, allowing us to neglect $\Phi$ when inverting. The drift-kinetic limit is particularly appropriate as we only treat the electron mirror terms implicitly while focusing on ion scale turbulence.

Repeating this exercise with $A_{\parallel}$ yields
\begin{align}
    H_{\ell,m+1}^e &= G_{\ell,m+1}^e + \frac{v_{te}}{\tau_e}\mathcal{J}_{\ell}^e A_{\parallel} \delta_{m+1,1}, \\
    H_{\ell-1,m+1}^e &= G_{\ell-1,m+1}^e + \frac{v_{te}}{\tau_e}\mathcal{J}_{\ell-1}^e A_{\parallel} \delta_{m+1,1}, \\
    H_{\ell,m-1}^e &= G_{\ell,m-1}^e + \frac{v_{te}}{\tau_e}\mathcal{J}_{\ell}^e A_{\parallel} \delta_{m-1,1}, \\
    H_{\ell+1,m-1}^e &= G_{\ell+1,m-1}^e + \frac{v_{te}}{\tau_e}\mathcal{J}_{\ell+1}^e A_{\parallel} \delta_{m-1,1}.
\end{align}
When taking a drift-kinetic limit and including prefactors, we only need include the $A_{\parallel}$ contribution to the $H_{\ell,m+1}^e$ and $H_{\ell-1,m+1}^e$ terms. Substituting in $A_{\parallel}$, the electron mirror term is then
\begin{equation}
\begin{split}
    v_{te}  \Big[& - (\ell+1) \, \sqrt{m+1} \, G^e_{\ell,m+1} - \ell \, \sqrt{m+1} \, G^e_{\ell-1,m+1}
+ \ell \, \sqrt{m} \, G^e_{\ell,m-1}
  \\&+ (\ell+1) \, \sqrt{m} \, G^e_{\ell+1,m-1}
     - \frac{v_{te}}{\tau_e}A_{\parallel}\delta_{\ell 0}\delta_{m 0} - \frac{v_{te}}{\tau_e}A_{\parallel}\delta_{\ell 1}\delta_{m 0} \Big] \nabla_\parallel \ln B.
\end{split}
\end{equation}

In $A_{\parallel}$ there are both ion and electron contributions $A_{\parallel} = A_{\parallel}^i + A_{\parallel}^e$. However, as we are only treating the electron mirror term implicitly, at each step of of our timestepping schemes, the ion term has already been advanced explicitly. Therefore, $A_{\parallel}^i$ can be moved to the RHS of the equation. We are only left with the electron contribution
\begin{equation}
\begin{split}
    v_{te}  \Big[& - (\ell+1) \, \sqrt{m+1} \, G^e_{\ell,m+1} - \ell \, \sqrt{m+1} \, G^e_{\ell-1,m+1}
+ \ell \, \sqrt{m} \, G^e_{\ell,m-1}
  \\&+ (\ell+1) \, \sqrt{m} \, G^e_{\ell+1,m-1}
     - \frac{v_{te}}{\tau_e}A_{\parallel}^e\delta_{\ell 0}\delta_{m 0} - \frac{v_{te}}{\tau_e}A_{\parallel}^e\delta_{\ell 1}\delta_{m 0} \Big] \nabla_\parallel \ln B.
\end{split}
\end{equation}

Finally, the equation we would solve is 
\begin{equation}
    \left(\mathsfbi{B}_{\mathrm{band}} + \boldsymbol{u}_3 \otimes \boldsymbol{v}_3\right)\boldsymbol{G}^e = \boldsymbol{r},
\end{equation}
where $\mathsfbi{B}_{\mathrm{band}}$ is a banded matrix formed from the coefficients of $G^e_{\ell,m}$, $G^e_{\ell,m+1}$, $G^e_{\ell,m-1}$, $G^e_{\ell-1,m+1}$, and $G^e_{\ell+1,m-1}$, and $\boldsymbol{u}_3$ and $\boldsymbol{v}_3$ are
% \begin{gather}
%     \boldsymbol{u}_3 = \left[-\frac{Z_iv_{ti}}{\tau_i}\delta_{\ell,0}\delta_{m,0} \ \ \boldsymbol{0} \ -\frac{Z_iv_{ti}}{\tau_i}\delta_{\ell,1}\delta_{m,0} \ \ \boldsymbol{0} \ \right. 
%     \left. -\frac{Z_ev_{te}}{\tau_e}\delta_{\ell,0}\delta_{m,0} \ \ \boldsymbol{0} \ -\frac{Z_ev_{te}}{\tau_e}\delta_{\ell,1}\delta_{m,2} \ \boldsymbol{0}\right] \label{mirror apar u}, \\
%     \boldsymbol{v}_3 = \left[0 \ \ \frac{\beta_{ref}}{2F}Z_in_iv_{ti}\delta_{\ell,0}\delta_{m,1} \ \ \boldsymbol{0} \ \ \frac{\beta_{ref}}{2F}Z_en_ev_{te}\delta_{\ell,0}\delta_{m,1} \ \ \boldsymbol{0}\right]\label{mirror apar v}.
% \end{gather}
\begin{gather}
    \boldsymbol{u}_3 = -v_{te}\nabla_\parallel \ln B \Delta t\left[\frac{v_{te}}{\tau_e}\delta_{\ell 0}\delta_{m 0} \ \ \boldsymbol{0} \ \frac{v_{te}}{\tau_e}\delta_{\ell 1}\delta_{m 0} \ \boldsymbol{0}\right] \label{mirror apar u},
\end{gather}
and
\begin{gather}
    \boldsymbol{v}_3 = \left[0 \ -\frac{\beta_{\mathrm{ref}}}{2F}n_ev_{te}\delta_{\ell 0}\delta_{m 1} \ \ \boldsymbol{0}\right]\label{mirror apar v}.
\end{gather}

\subsection{Hypercollisions}\label{invert hypercollisions}
Since the parallel streaming terms are Fourier transformed in {\tt GX}, there is no inherent parallel dissipation that would normally arise from upwinding in other gyrokinetic codes. Furthermore, in the absence of sophisticated closures, additional dissipation is needed in velocity space to dampen the higher $\ell$ or $m$ modes. As a result, {\tt GX} is normally run with a hypercollision operator which is derived in \cite{Mandell2024GX:Design}. This operator takes the form
\begin{equation}
\left(\pderiv{G_{\ell,m}}{t}\right)_\textrm{hyp-coll}  =
%\notag \\ &\quad
-\nu_\mathrm{hyp} m^{p} G_{\ell, m} \quad (m>2),
\end{equation}
where the collision frequency $\nu_\mathrm{hyp}$ is
\begin{equation}
     \nu_\mathrm{hyp} = 2.5 f_\mathrm{hyp} \frac{p+1/2}{m_\mathrm{max}^{p+1/2}}|k_\parallel|v_{ts}.
\end{equation}
Due to the factor of $v_{ts}$, this term can become stiff for electrons. However, this term is straightforward to handle alongside the parallel streaming terms, as it is also evaluated spectrally in $z$. Since this term is only on the main diagonal, it can be directly included in the tridiagonal solver routine.

\subsection{Remaining Terms}\label{remaining terms}
The remaining terms that are not evolved implicitly  correspond physically to the magnetic and curvature drifts, drive terms from the temperature and density gradients, and the collision operator. Although all of these terms are usually about the same order of magnitude between ions and electrons, the magnetic and curvature drift terms can become stiff due to the prefactor of $\omega_d^{\kappa}$ or $\omega_d^{\bnabla B}$ which are
\begin{gather}
    \mathrm{i} \omega_d^{\kappa} = (1/B)\bhat\times(\bhat\bcdot\bnabla\bhat)\bcdot {\mathrm{i}\boldsymbol k_\perp},
\end{gather}
and
\begin{gather}
    \mathrm{i} \omega_d^{\bnabla B} = (1/B^2)\bhat\times\bnabla B\bcdot {\mathrm{i}\boldsymbol k_\perp}.
\end{gather}
If the geometry of the fusion device has particularly strong curvature or $\bnabla B$, or if one wants to investigate electron scale turbulence, the drift terms can dominate the CFL condition instead of the streaming or mirror terms. Indeed, we show in the next section that these drift terms will limit the maximum stable timestep for the tokamak cases. In practice, it is straightforward to treat these terms implicitly together with the mirror terms. However, the $k_{\perp}$ dependence of $\omega_d^{\kappa}$ and $\omega_d^{\bnabla B}$ results in $N_x \, N_y \, N_z$ matrices that would need to be stored (as opposed to just $N_z$ mirror matrices). This memory cost is prohibitively expensive on a single GPU, even for banded matrices. While we present some preliminary results  of implicit evaluation of these terms in Appendix \ref{implicit drifts examples}, we leave a full implementation to future work.

\subsection{Stationary Iteration}\label{stationary iteration}

The approximations introduced in \S\ref{invert stream} and \S\ref{invert stream with apar} to invert the streaming operator can introduce significant errors that degrade both the accuracy and stability of the numerical method. To address these issues, we apply a simple stationary iteration to improve the solution at each implicit step.

Specifically, at each IMEX timestep, we solve a linear system of the form
\begin{equation}
    \mathsfbi{L}\boldsymbol{G} = \boldsymbol{r},
\end{equation}
where $\mathsfbi{L}$ is the linear operator and $\boldsymbol{r}$ is the right-hand side. Rather than inverting $\mathsfbi{L}$ directly, we introduce a preconditioner $\mathsfbi{P} \approx \mathsfbi{L}$, and rewrite the equation as
\begin{equation}
    (\mathsfbi{L} - \mathsfbi{P} + \mathsfbi{P})\boldsymbol{G} = \boldsymbol{r},
\end{equation}
which can be rearranged into a fixed-point form:
\begin{equation}
    \mathsfbi{P}\boldsymbol{G} = (\mathsfbi{P} - \mathsfbi{L})\boldsymbol{G} + \boldsymbol{r}.
\end{equation}
This motivates the following iterative update (where the subscript index indicates the iteration step number, not the time index):
\begin{equation}
    \boldsymbol{G}_{n+1} = (\mathsfbi{I} - \mathsfbi{P}^{-1}\mathsfbi{L})\boldsymbol{G}_n + \mathsfbi{P}^{-1}\boldsymbol{r}.
\end{equation}
This scheme converges when the spectral radius of $\mathsfbi{I} - \mathsfbi{P}^{-1}\mathsfbi{L}$ is less than one. When the preconditioner $\mathsfbi{P}$ is a good approximation to $\mathsfbi{L}$, then $ \mathsfbi{P}^{-1}\mathsfbi{L} \approx \mathsfbi{I} $, and convergence is typically rapid. In the cases of \S\ref{invert stream} and \S\ref{invert stream with apar}, $\mathsfbi{B}$ is the linear operator, and the derived simplified matrices represent the preconditioner.

While this approach is simple and effective, more advanced iterative solvers such as GMRES or BiCGSTAB may provide more robust convergence. These methods, along with a suite of IMEX integrators, are available in the {\tt SUNDIALS} library \citep{Hindmarsh2005SUNDIALS:Solvers, Balos2021EnablingLibrary, Gardner2022EnablingSolvers, Reynolds2023ARKODE:Methods}. Interfacing {\tt GX} with {\tt SUNDIALS} will be the subject of future work. Importantly, if the current algorithm converges using basic stationary iteration, significant performance improvements can be expected by switching to Krylov methods.

\subsection{Operator-Split Preconditioning}\label{operator split preconditioning}

The ARK methods used in this work require that the parallel streaming and magnetic mirror terms be combined and inverted together. However, the magnetic mirror terms are not Fourier-transformed and thus cannot be easily combined with the parallel streaming operator into a single matrix. To address this, we apply an operator-splitting approximation.

Ideally, we would invert a matrix of the form
\begin{equation}
    \mathsfbi{I} - \Delta t \mathsfbi{B} - \Delta t \mathsfbi{C},
\end{equation}
where $\mathsfbi{B} $ represents the parallel streaming operator and $\mathsfbi{C} $ represents the mirror force operator (or similar). Rather than constructing and inverting the combined matrix, we approximate the inverse by sequentially applying each operator:
\begin{equation}
    (\mathsfbi{I} - \Delta t \mathsfbi{B})(\mathsfbi{I} - \Delta t \mathsfbi{C}) = \mathsfbi{I} - \Delta t \mathsfbi{B} - \Delta t \mathsfbi{C} + \left(\Delta t\right)^2 \mathsfbi{B}\mathsfbi{C}.
\end{equation}
This approximation introduces a local error of second order in $\Delta t$ due to the $\left(\Delta t\right)^2 \mathsfbi{B}\mathsfbi{C}$ term. To reduce this error, we employ an `outer' stationary iteration that simultaneously includes both the streaming and mirror operators in the iteration loop, using the operator split approximation as the preconditioner.

For the examples shown in this paper, we find that this combined iteration eliminates the need for a separate stationary iteration for the streaming term alone, thereby providing a simple approximation for the full implicit solve.

\section{Numerical Results}\label{numerical results}
In this section, we present results from several simulations that test our implicit methods. The simulations in \S\ref{slab results} are conducted in slab geometry to test the implicit treatment of the parallel streaming terms. The simulations in \S\ref{cbc results} are performed in toroidal geometry to test the combined inversion of the parallel streaming and mirror terms. We initially focus on evaluating the accuracy of the implicit methods before giving a more detailed discussion of the timings in \S\ref{timings section}.
\subsection{Slab Geometry}\label{slab results}
To test our preconditioner for the parallel streaming terms, we start in slab geometry, where the mirror and drift terms are zero. The geometry also includes finite magnetic shear so that the perpendicular wavenumbers $k_{\perp}$ vary along the field-line.

\begin{figure}
     \centering
     \begin{subfigure}[b]{0.48\textwidth}
         \centering         \includegraphics[width=\textwidth]{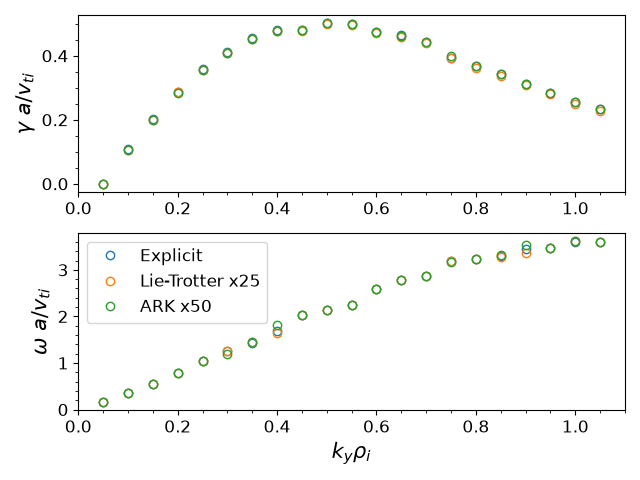}
         \caption{}
         \label{linear slab es}
     \end{subfigure}
     \hfill
     \begin{subfigure}[b]{0.48\textwidth}
         \centering
         \includegraphics[width=\textwidth]{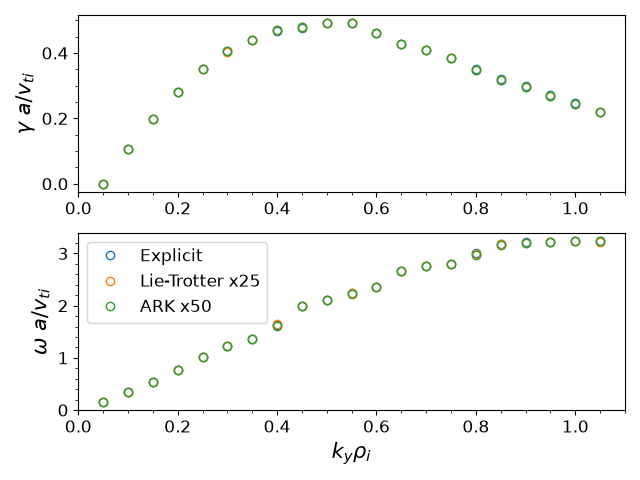}
         \caption{}
         \label{linear slab em}
     \end{subfigure}
     \hfill
        \caption{The linear growth rates and frequencies for the explicit RK3 (blue), the Lie-Trotter (orange), and the ARK (green) methods in slab geometry. The left plot is for the electrostatic case, while the right plot is at $\beta = 4\%$. The `x' indicates the ratio of the IMEX timestep to the explicit timestep.}
        \label{linear slab}
\end{figure}

The results in figure \ref{linear slab es} show the linear growth rates and frequencies in this slab geometry. The magnetic shear is $\hat{s} = (1/q)\,\mathrm{d}q/\mathrm{d}\psi = 0.8$, which is comparable to shear in the core of a tokamak. The simulations using the Lie-Trotter and ARK methods employ a timestep of 25 and 50 times larger, respectively, than that used in the explicit scheme. As seen in the plot, the agreement between all three methods is excellent. Although the Lie-Trotter method is numerically stable with a 50 times greater timestep, there is worse agreement.

To test the electromagnetic implicit scheme, we set $\beta = 4\%$ and reran the simulations with the same magnetic geometry. The results are shown in figure \ref{linear slab em}. As in the electrostatic case, the IMEX methods achieve good accuracy even when including electromagnetic fluctuations.

% \begin{figure}
%      \centering
%      \begin{subfigure}[b]{0.48\textwidth}
%          \centering         \includegraphics[width=\textwidth]{Figures/slab_es_phi2_ky0_4.png}
%          \caption{}
%          \label{linear slab es}
%      \end{subfigure}
%      \hfill
%      \begin{subfigure}[b]{0.48\textwidth}
%          \centering
%          \includegraphics[width=\textwidth]{Figures/slab_em_phi2_ky0_6.png}
%          \caption{}
%          \label{linear slab em}
%      \end{subfigure}
%      \hfill
%         \caption{Plots of the electrostatic potential $|\Phi^2|$ versus time at $k_y = 0.4$ and $k_y = 0.6$ for the two cases in Fig. \ref{linear slab}. Despite discrepancies on the plotted growth rates, the evolution of potential closely matches.}
%         \label{linear slab phi2}
% \end{figure}

\begin{figure}
    \centering
    \includegraphics[width=0.5\linewidth]{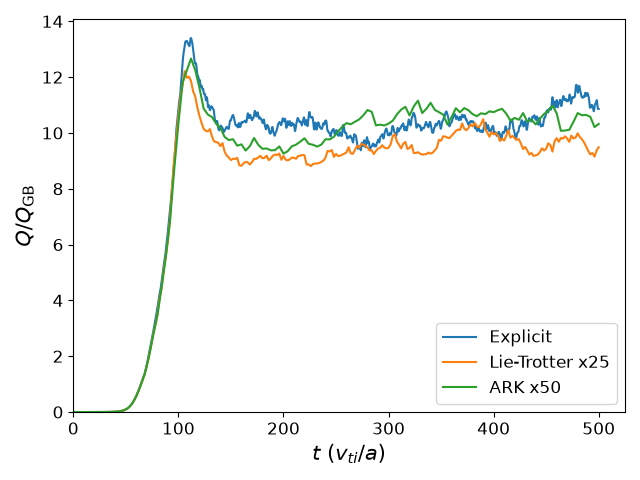}
    \caption{The ion heat flux for the nonlinear electrostatic slab case between the explicit RK3 (blue), Lie-Trotter (orange), and ARK (green) methods. The electron heat flux in this case is negligible, and so not plotted. The `x' indicates the ratio of the IMEX timestep to the explicit timestep.}
    \label{nonlinear slab}
\end{figure}

Finally, we run an electrostatic nonlinear simulation, and plot the resulting turbulent ion heat fluxes in figure \ref{nonlinear slab}. The plot shows excellent agreement in the heat fluxes between the RK3 and ARK methods. However, even at a reduced timestep of only 25 times the explicit method, the Lie-Trotter method does not recover a similar average heat flux.

\subsection{Tokamak Geometry}\label{cbc results}
In this section, we compare the different schemes in tokamak-like geometry. The geometry we use is the `Cyclone Base Case' (CBC), which is commonly used to benchmark gyrokinetic codes \citep{Dimits2000ComparisonsSimulations}.

\begin{figure}
     \centering
     \begin{subfigure}[b]{0.48\textwidth}
         \centering
         \includegraphics[width=\textwidth]{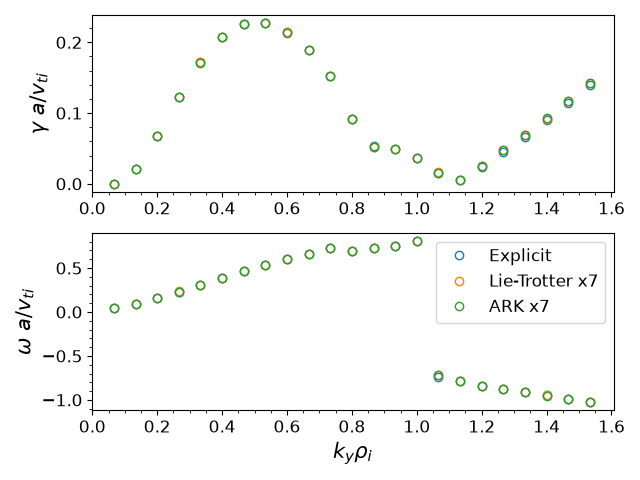}
         \caption{}
         \label{linear cbc es}
     \end{subfigure}
     \hfill
     \begin{subfigure}[b]{0.48\textwidth}
         \centering
         \includegraphics[width=\textwidth]{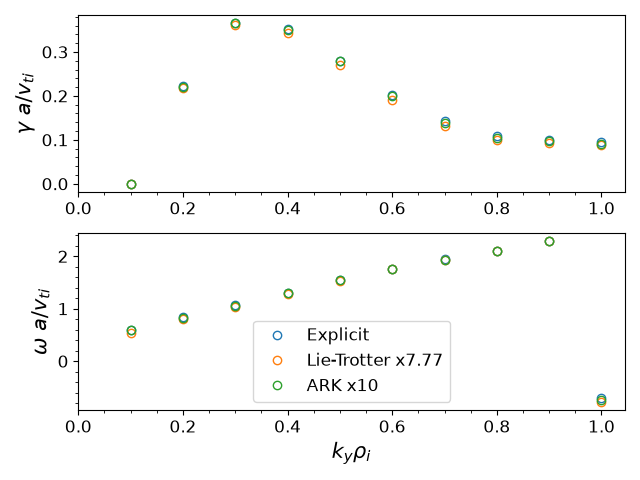}
         \caption{}
         \label{linear cbc em}
     \end{subfigure}
     \hfill
        \caption{The linear growth rates and frequencies for the explicit RK3 (blue), the Lie-Trotter (orange), and the ARK (green) methods for the cyclone base-case. The left plot is for an electrostatic simulation, while the right is at $\beta = 1.5\%$. The `x' indicates the ratio of the IMEX timestep to the explicit timestep.}
        \label{linear cbc}
\end{figure}

The results in figure \ref{linear cbc} show the linear growth rates and frequencies for each simulation. Unlike in slab geometry, in tokamak geometry the implicit methods cannot use a 60 times larger timestep than that of the explicit method. As will be explained in \S\ref{timings section}, the magnetic and curvature drift terms dominate the CFL constraint for the implicit methods. In this case, we can only use a 7--10 times greater timestep. Nevertheless, at this timestep, we achieve good agreement between all of the schemes for both the electrostatic and electromagnetic simulations.

\begin{figure}
    \centering
    \includegraphics[width=0.5\linewidth]{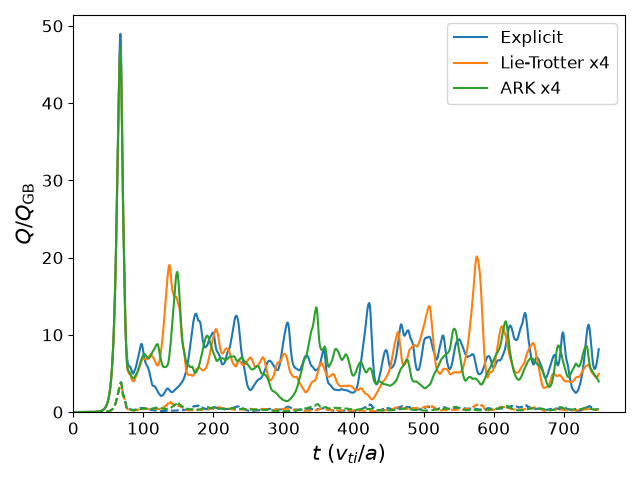}
    \caption{The ion (solid) and electron (dashed) heat fluxes for the electrostatic cyclone base-case for the explicit RK3 (blue), the Lie-Trotter (orange), and the ARK (green) methods. The `x' indicates the ratio of the IMEX timestep to the explicit timestep.}
    \label{cbc nonlinear}
\end{figure}

As a final test, we run nonlinear electrostatic simulations of the CBC. In this case, due to the magnetic/curvature drifts and the nonlinear term, the IMEX methods can only take up to a 4 times larger timestep. The resulting heat fluxes are shown in figure \ref{cbc nonlinear}. As seen in the plot, there is good agreement in the heat fluxes between the methods.

\subsection{Convergence of the Iterative Method for the Streaming Terms}
\begin{figure}
     \centering
     \begin{subfigure}[b]{0.48\textwidth}
         \centering
         \includegraphics[width=\textwidth]{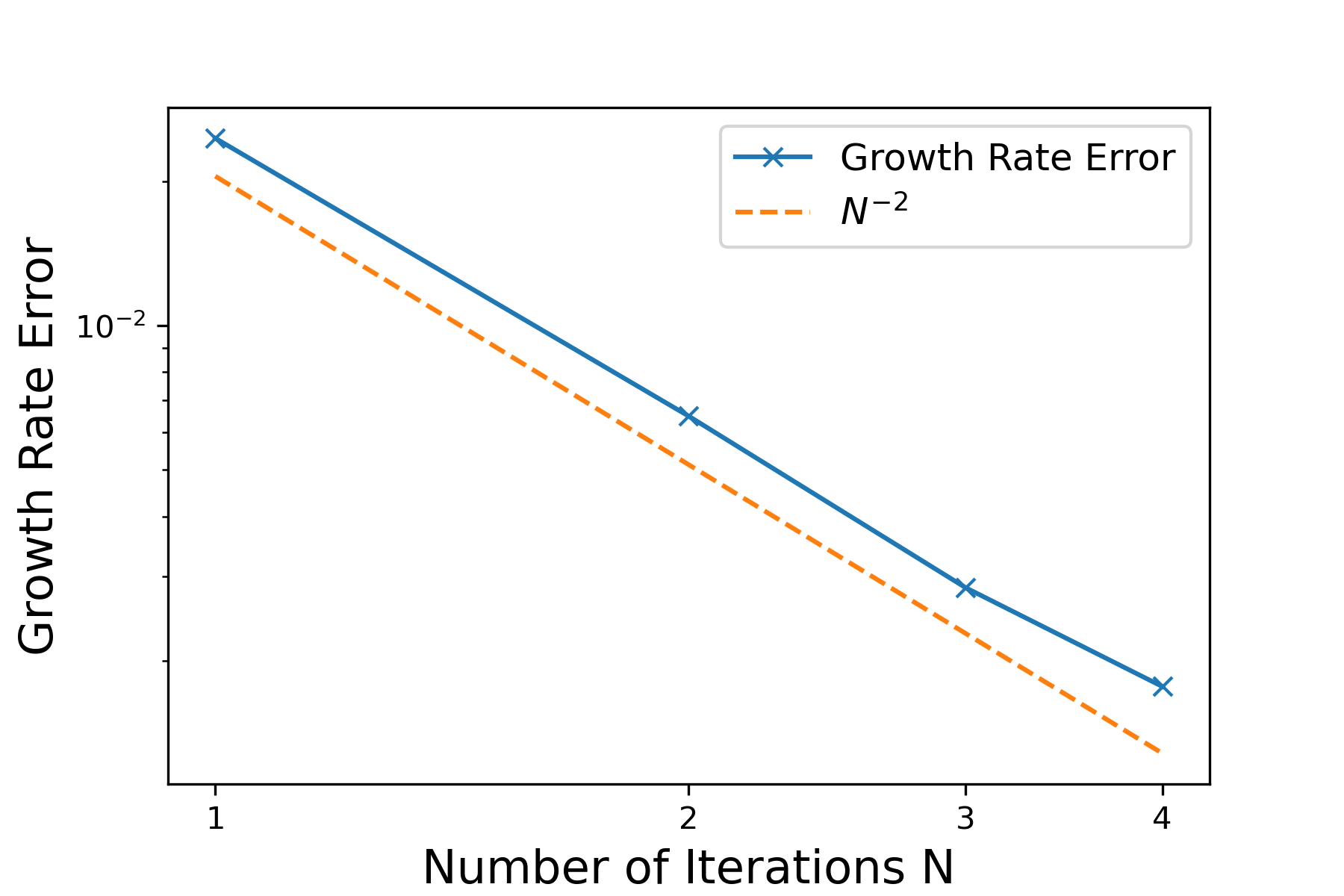}
         \caption{}
         \label{linear slab convergence}
     \end{subfigure}
     \hfill
     \begin{subfigure}[b]{0.48\textwidth}
         \centering
         \includegraphics[width=\textwidth]{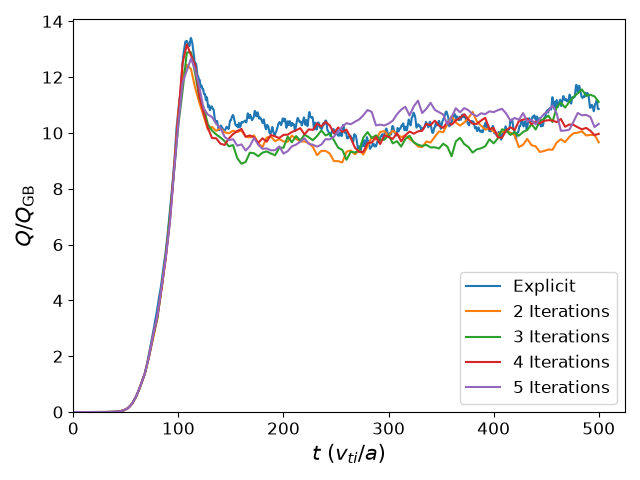}
         \caption{}
         \label{nonlinear slab convergence}
     \end{subfigure}
     \hfill
        \caption{(a) The growth rate error at $k_y = 0.4$ versus the number of iterations for the iterative method for the streaming terms. The blue curve is the actual error, while the orange dashed line is $N^{-2}$ that the error seems to follow. (b) The nonlinear heat fluxes for the ARK method for different number of iterations. The average heat flux eventually matches that from the explicit method at 5 iterations.}
        \label{slab convergence}
\end{figure}

To analyse the convergence of the iterative method used to invert the parallel streaming operator, we perform the electrostatic linear and nonlinear slab simulations with different numbers of stationary iterations using the ARK method. Figure \ref{linear slab convergence} shows the relative error in the growth rate at $k_y = 0.4$ versus the number of iterations. The errors decrease rapidly, reaching about a 0.1\% error at 4 iterations. The decrease in the error follows approximately a $1/N^2$ scaling, where $N$ is the number of iterations. Figure \ref{nonlinear slab convergence} shows the nonlinear heat flux traces for different numbers of iterations. We observe the steady-state heat flux from the implicit simulations consistently converges towards the heat flux from the explicit simulation as the number of iterations increases. Overall, this demonstrates that our simple preconditioned iterative solver is effective, and could have potential for further improvement using more sophisticated iterative methods.

\subsection{Timings}\label{timings section}
Table \ref{tab:timing-summary} shows the timings of the different simulations shown in the previous sections.  For the slab cases, the Lie-Trotter and ARK methods maintain a speedup of over an order of magnitude. This speedup reduces in the CBC cases in part due to the CFL condition being set by the explicit drift terms. Following the procedure in \citet{Mandell2024GX:Design}, we expect the CFL condition to be set by a combination of the drift and ion mirror terms
\begin{equation}
    \omega_{\mathrm{max}} = \omega_{\mathrm{drifts,max}} + \omega_{\mathrm{ion, mirror,max}}.
\end{equation}

The maximum drift frequency can be estimated as
\begin{equation}
    \omega_{\mathrm{drifts,max}} \sim \omega_{d,\mathrm{max}}^{\kappa} v_{\parallel,\mathrm{max}}^2 + \omega_{d,\mathrm{max}}^{\bnabla B} \left(\mu B\right)_{\mathrm{max}},
\end{equation}
where $v_{\parallel,\mathrm{max}}$ and $\left(\mu B\right)_{\mathrm{max}}$ scale like
\begin{gather}
    v_{\parallel,\mathrm{max}} \sim 2 \mathrm{He}_{\mathrm{c},\mathrm{max}},
\end{gather}
and
\begin{gather}
    \left(\mu B\right)_{\mathrm{max}} \sim \mathrm{L}_{\mathrm{c},\mathrm{max}},
\end{gather}
and $\mathrm{He}_{\mathrm{c},\mathrm{max}}$ and $\mathrm{L}_{\mathrm{c},\mathrm{max}}$ are the maximum values of the Gauss-Hermite and Gauss-Laguerre collocation points. For the linear CBC simulations in this work, $\omega_{\mathrm{drifts,max}} \approx 200k_{\perp}$, while $\omega_{\mathrm{ion, mirror,max}} \approx 1.5$. Since $\omega_{\mathrm{drifts,max}}$ scales with the perpendicular wavenumber $k_{\perp}$, the maximum timestep in the electromagnetic CBC simulation was slightly larger than in the electrostatic simulation, as it had a smaller maximum $k_y$. For $k_y \approx 1$, we expect a maximum stable timestep of about $1/\omega_{\mathrm{max}} \approx \num{5.0e-3}$ $v_{ti}/a$.

In comparison, the streaming terms would contribute a term that scales like
\begin{equation}
    \omega_{\mathrm{stream,max}} \sim \left(\bhat \bcdot \bnabla z\right)k_{z,\mathrm{max}}v_{\parallel,\mathrm{max}}v_{ts}.
\end{equation}
For electrons, this term is approximately $\omega_{\mathrm{stream,max}} \approx 1980$, making the maximum timestep $1/\omega_{\mathrm{stream,max}} \approx \num{5.0e-4}$ $v_{ti}/a$. Therefore, we expect that the implicit methods can only use about a 10 times greater timestep than the explicit method, which is demonstrated in the CBC simulations.

\begin{figure}
     \centering
     \begin{subfigure}[b]{0.48\textwidth}
         \centering
         \includegraphics[width=\textwidth]{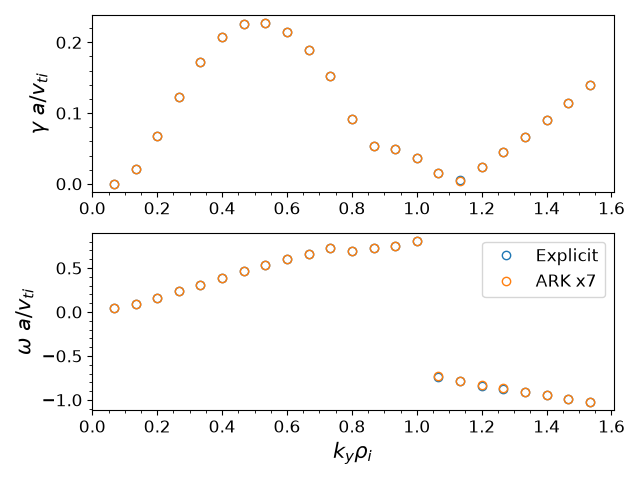}
         \caption{}
         \label{linear cbc es exp mirror}
     \end{subfigure}
     \hfill
     \begin{subfigure}[b]{0.48\textwidth}
         \centering
         \includegraphics[width=\textwidth]{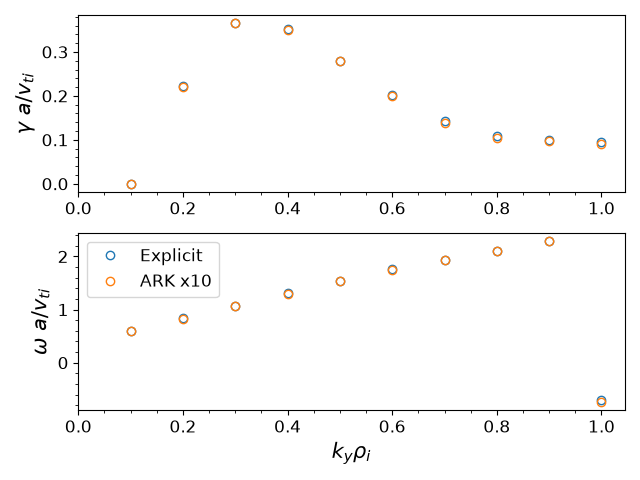}
         \caption{}
         \label{linear cbc em exp mirror}
     \end{subfigure}
     \hfill
        \caption{The linear growth rates and frequencies for the explicit RK3 (blue) and the ARK methods for the cyclone base-case. The left plot is for an electrostatic simulation, while the right is at $\beta = 1.5\%$. The `x' indicates the ratio of the IMEX timestep to the explicit timestep. In these cases, the ARK method evaluates the mirror term explicitly.}
        \label{linear cbc explicit mirror}
\end{figure}

Finally, we find that the most expensive operation in our IMEX schemes is inverting the matrices for the mirror operator. This cost increases as we increase the number of iterations, particularly for the ARK scheme. To demonstrate this effect, we run the CBC simulations again with the ARK method, but treat the mirror term explicitly. As shown in figure \ref{linear cbc explicit mirror} and table \ref{tab:timing-summary}, we can achieve comparable accuracy and maximum stable timestep at a reduced wall-time. Therefore, future work will investigate different sparse linear algebra libraries to reduce this cost. Nevertheless, for the nonlinear case, we can still achieve approximately a factor of 2 speedup.

\begin{table}
\centering
\begin{tabular}{@{} l S[table-format=1.1e-2] S[table-format=2.1] S[table-format=3.2] S[table-format=2.2] @{}}
\toprule
Scheme & {$\Delta t$ ($v_{ti}/a$)} & {$\Delta t$ Factor} & {Wall Time (min)} & {Speedup} \\
\midrule
\multicolumn{5}{l}{\textbf{Sheared Slab (ES)}} \\
RK3 & \num{8.0e-4} & 1 & 8.68 & 1.00 \\
Lie-Trotter (3 stream iters) & \num{2.0e-2} & 25 & 0.79 & 10.99 \\
ARK (3 stream iters) & \num{4.0e-2} & 50 & 0.52 & 16.69 \\
\addlinespace

\multicolumn{5}{l}{\textbf{Sheared Slab (EM)}} \\
RK3 & \num{8.0e-4} & 1 & 8.19 & 1.00 \\
Lie-Trotter (3 stream iters) & \num{2.0e-2} & 25 & 0.96 & 8.53 \\
ARK (3 stream iters) & \num{4.0e-2} & 50 & 0.63 & 13.0 \\
\addlinespace

\multicolumn{5}{l}{\textbf{Sheared Slab (Nonlinear, ES)}} \\
RK3 & \num{8.0e-4} & 1 & 489.49 & 1.00 \\
Lie-Trotter (8 stream iters) & \num{2.0e-2} & 25 & 39.88 & 12.27 \\
ARK (5 stream iters) & \num{4.0e-2} & 50 & 23.56 & 20.78 \\
\addlinespace

\multicolumn{5}{l}{\textbf{CBC (ES)}} \\
RK3 & \num{7.0e-4} & 1 & 5.85 & 1.00 \\
Lie-Trotter (2 stream iters) & \num{5.0e-3} & 7.14 & 1.49 & 3.93 \\
ARK (1 outer iters) & \num{5.0e-3} & 7.14 & 1.57 & 3.73 \\
ARK (1 stream iters, explicit mirror) & \num{5.0e-3} & 7.14 & 1.25 & 4.68 \\
Lie-Trotter (4 stream iters, implicit drifts) & \num{1.0e-3} & 14.29 & 1.05 & 5.57 \\
\addlinespace

\multicolumn{5}{l}{\textbf{CBC (EM)}} \\
RK3 & \num{9.0e-4} & 1 & 2.56 & 1.00 \\
Lie-Trotter (3 stream iters) & \num{7.0e-3} & 7.78 & 1.12 & 2.29 \\
ARK (2 outer iters) & \num{9.0e-3} & 10.0 & 1.24 & 2.06 \\
ARK (2 stream iters, explicit mirror) & \num{9.0e-3} & 10.0 & 0.69 & 3.71 \\
\addlinespace

\multicolumn{5}{l}{\textbf{CBC (Nonlinear, ES)}} \\
RK3 & \num{1.0e-3} & 1 & 617.97 & 1.00 \\
Lie-Trotter (3 stream iters) & \num{4.0e-3} & 4.0 & 217.83 & 2.84 \\
ARK (2 outer iters) & \num{4.0e-3} & 4.0 & 331.58 & 1.86 \\
\bottomrule
\end{tabular}
\caption{Simulation timings for each scheme and configuration. ES and EM denote electrostatic and electromagnetic simulations, respectively.}
\label{tab:timing-summary}
\end{table}

\section{Conclusions}\label{conclusions}
In this work, we developed implicit-explicit methods to accelerate kinetic-electron simulations for the {\tt GX} gyrokinetics code. Since {\tt GX} is GPU-native, we have designed our methods to be low-memory by taking advantage of banded and low-rank structure of the {\tt GX} equations to efficiently invert the implicit operators. We showed that our methods can achieve up to the maximum theoretical timestep while still maintaining an improvement in the wall-clock time for both linear and nonlinear simulations. However, this scheme uses approximate iterative methods. While we show that the iterative method for the streaming terms is efficient and accurate, the operator-split preconditioned iterative method can be more expensive due to the cost of direct banded solvers. These methods are also less effective in cases where the drift terms dominate the CFL condition.

There are several directions for possible future work. We first aim to implement these methods within the {\tt SUNDIALS} framework. Doing so should improve the relatively simple iterative methods shown here, and improve the efficiency of the ARK methods. The {\tt SUNDIALS} library also includes robust methods for error-checking and adaptive timesteps. Additionally, we will include a full implementation of the magnetic and curvature drift terms in the implicit operator, which would further relax the CFL constraint.

\section*{Acknowledgments}
P. S. K. would like to acknowledge F. Parra Diaz for several helpful conversations.  This work was supported in part by the U.S. Department of Energy, Office of Science, Office of Advanced Scientific Computing Research, Scientific Discovery through Advanced Computing (SciDAC) Program through the Computational Evaluation and Design of Actuators for Core-Edge Integration (CEDA) project, and through the Frameworks, Algorithms and Software Technologies for Mathematics (FASTMath) Institute. This material is based upon work supported by the U.S. Department of
Energy, Office of Science, Office of Advanced Scientific Computing Research, Department of Energy Computational Science Graduate Fellowship under Award Number DE-SC0024386.

P. S. K. used ChatGPT (OpenAI) and Claude (Anthropic) to aid in software development and debugging and in revising text. P. S. K reviewed any AI-assisted code and text, and the manuscript has been reviewed by the authors, who take full responsibility for its content.
\appendix

\section{Normalizations}\label{normalizations}
The normalizations used in this work are the same as in \citet{Mandell2024GX:Design} and are summarized in tables \ref{tab:Table-independent-norm} and \ref{tab:Table-dependent-norm}. The normalizing quantities in table \ref{tab:Table-independent-norm} are arbitrary, but have common choices. For example, the normalizing length $a_N$ is typically either the minor or major radius. The reference magnetic field $B_N$ is typically taken to be the vacuum magnetic field at the magnetic axis or is defined in terms of the enclosed toroidal magnetic flux. The reference thermal speed $v_{\mathrm{th,ref}}$ is defined in terms of a reference temperature and mass $v_{\mathrm{th,ref}} = \sqrt{T_{\mathrm{ref}}/M_{\mathrm{ref}}}$. The corresponding reference gyroradius and cyclotron frequency are then $\rho_{\rm{ref}}  \equiv v_{\rm{th, ref}}/\Omega_{\rm ref}$ and $\Omega_{\rm{ref}} = q_{\rm{ref}} B_{\rm{N}}/M_{\rm{ref}} c$, respectively. Specific species quantities such as the temperature, mass, thermal speed, charge, and gyroradius are then normalized with respect to these reference values: $\tau_s = T_s/T_\mathrm{ref}$, $m_s = M_s/M_\mathrm{ref}$, $v_{ts} = v_{\mathrm{th},s}/v_\mathrm{th,ref}$, $Z_s = q_s/q_\mathrm{ref}$, and $\rho_s = (v_{\mathrm{th},s}/\Omega_s)/\rho_\mathrm{ref}$.

\begin{table}
\centering
\begin{tabular}{c|c|c|c|c|c}
 $k_x, k_y, x, y$ & $\boldsymbol{\bnabla}, z$ & $v_\parallel$ & $\mu$ & $\omega, t$  & $\psi$ \\[3pt]
 $\rho_{\rm{ref}}$  & $a_{\rm{N}}$ & $v_{\rm{th},s}$ & $v_{\rm{th},s}^2{\color{black}/B_\mathrm{N}}$  & $v_{\rm{th, ref}}/a_{\rm{N}}$ & $a_{\rm{N}}^2B_{\rm{N}}$\\
\end{tabular}
\caption{Normalizing the quantities for the independent variables. Taken from \citet{Mandell2024GX:Design}}
\label{tab:Table-independent-norm}
\end{table}
\begin{table}
\centering
\begin{tabular}{c|c|c|c}
 $h$ & $\Phi$ & $A_{\parallel}$ & $\delta\!B_{\parallel}$  \\[3pt]
 $\rho_* n_0/v^3_{\rm{th, ref}}$  & $\rho_{*}T_{\rm{ref}}/q_{\rm{ref}}$  & $\rho_{*}(\rho_\mathrm{ref} B_{\rm{N}})$ & $\rho_{*} B_{\rm{N}}$\\
\end{tabular}
\caption{Normalizing quantities for dependent variables. Taken from \citep{Mandell2024GX:Design}}
\label{tab:Table-dependent-norm}
\end{table}
\section{Long-Wavelength Dispersion Relation In Slab Geometry}\label{app a}

In unsheared slab geometry, we consider a simplified limit with no gradients or collisions, stationary ions, and isothermal electrons. In the long-wavelength limit, the {\tt GX} equations for the electron moments reduce to
\begin{gather}
    \frac{\partial G_0^e}{\partial t} +  v_{te}\nabla_{\parallel}\left(G_1^e + A_{\parallel}\right) = 0,
\end{gather}
and
\begin{gather}
    \frac{\partial G_1^e}{\partial t} + v_{te}\nabla_{\parallel}\left(G_0^e - \Phi\right) = 0,
\end{gather}
with the corresponding field equations given by
\begin{gather}
    \Phi = -\frac{G_0^e}{\left(k_{\perp}\rho_i\right)^2},
\end{gather}
and
\begin{gather}
    A_{\parallel} = -\frac{\hat{\beta}v_{te}G_1^e}{k_{\perp}^2 +\hat{\beta}},
\end{gather}
where we define $\hat{\beta} = \beta m_\mathrm{i}/2m_\mathrm{e}$.

Fourier transforming and combining the moment and field equations yields the dispersion relation:
\begin{equation}
    \omega^2 = v_{te}^2k_{\parallel}^2\left(1 + \frac{1}{k_{\perp}^2}\right) \left(1 - \frac{\hat{\beta}}{k_{\perp}^2 + \hat{\beta}}\right).
\end{equation}
This simplifies to the standard dispersion relation for the kinetic Alfvén wave:
\begin{equation}
    \omega^2 = v_{te}^2k_{\parallel}^2\left(\frac{1 + k_{\perp}^2}{k_{\perp}^2 + \hat{\beta}}\right).
\end{equation}
In the electrostatic limit (\(\hat{\beta} = 0\)), this further reduces to the \(\omega_H\) mode:
\begin{equation}
    \omega_H^2 = v_{te}^2 k_{\parallel}^2\frac{1 + k_{\perp}^2}{k_{\perp}^2}.
\end{equation}

This expression shows that the mode frequency increases significantly as \(k_{\perp} \to 0\). Therefore, to account for the fastest waves, we take the smallest \(k_{\perp}\) along the field line in constructing our preconditioner. Notably, explicit gyrokinetic codes often retain a small but nonzero \(\beta\) even for nominally electrostatic simulations in order to avoid this high-frequency mode.

For more general geometry and nonzero \(\hat{\beta}\), we approximate the dispersion relation as
\begin{equation}
    \omega^2 = v_{te}^2k_{\parallel}^2\left(1 + \frac{1}{\left(k_{\perp}\right)_m^2}\right) \left(1 - \frac{\hat{\beta}}{\left(k_{\perp}^2 + \hat{\beta}\right)_M}\right),
\end{equation}
where \((k_{\perp}^2)_m\) and \((k_{\perp}^2 + \hat{\beta})_M\) represent specific values chosen along the field line. Again, $k_{\perp}\rho$ is minimized, while $k_{\perp}^2 + \hat{\beta}$ is maximized to result in the maximum possible frequency

\section{Sherman-Morrison Formula}\label{app sherman morrison}
The matrices we invert in this work are of the form
\begin{equation}
    \left(\mathsfbi{A} + \boldsymbol{u} \otimes \boldsymbol{v}\right)\boldsymbol{G} = \boldsymbol{r},
\end{equation}
where $\boldsymbol{u}$ and $\boldsymbol{v}$ contain information about the fields. The Sherman-Morrison formula states that the solution to this equation is
\begin{equation}
    \boldsymbol{G} = \boldsymbol{y} - \left(\frac{\boldsymbol{v} \bcdot \boldsymbol{y}}{1 + \boldsymbol{v} \bcdot \boldsymbol{z}}\right) \boldsymbol{z},
\end{equation}
where $\boldsymbol{y}$ and $\boldsymbol{z}$ are solutions to the equations
\begin{equation}
    \mathsfbi{A}\boldsymbol{y} = \boldsymbol{r}, \quad  \mathsfbi{A}\boldsymbol{z} = \boldsymbol{u}.
\end{equation}

When including multiple fields, like $\Phi$ and $A_{\parallel}$, the equation becomes
\begin{equation}
    \left(\mathsfbi{A} + \boldsymbol{u}_1 \otimes \boldsymbol{v}_1 + \boldsymbol{u}_2 \otimes \boldsymbol{v}_2\right)\boldsymbol{G} = \boldsymbol{r}.
\end{equation}
Following the procedure in \citet{Tatsuno2008SecondaryInstability} and \citet{ Barnes2009LinearizedTests}, we rewrite the above equation as
\begin{equation}
    \left(\mathsfbi{A}_1 + \boldsymbol{u}_2 \otimes \boldsymbol{v}_2\right)\boldsymbol{G} = \boldsymbol{r},
\end{equation}
where $\mathsfbi{A}_1 = \mathsfbi{A} + \boldsymbol{u}_1 \otimes \boldsymbol{v}_1$. The solution is then
\begin{equation}
    \boldsymbol{G} = \boldsymbol{y}_2 - \left(\frac{\boldsymbol{v}_2 \bcdot \boldsymbol{y}_2}{1 + \boldsymbol{v}_2 \bcdot \boldsymbol{z}_2}\right) \boldsymbol{z}_2,
\end{equation}
where $\boldsymbol{y}_2$ and $\boldsymbol{z}_2$ are solutions to the equations
\begin{equation}
    \mathsfbi{A}_1\boldsymbol{y}_2 = \boldsymbol{r}, \quad  \mathsfbi{A}_1\boldsymbol{z}_2 = \boldsymbol{u}_2.
\end{equation}
To solve for $\boldsymbol{y}_2$ and $\boldsymbol{z}_2$, we apply the Sherman-Morrison formula a second time
\begin{gather}
    \boldsymbol{y}_2 = \boldsymbol{y}_1 - \left(\frac{\boldsymbol{v}_1 \bcdot \boldsymbol{y}_1}{1 + \boldsymbol{v}_1 \bcdot \boldsymbol{z}_1}\right) \boldsymbol{z}_1,
\end{gather}
and
\begin{gather}
    \boldsymbol{z}_2 = \boldsymbol{z}' - \left(\frac{\boldsymbol{v}_1 \bcdot \boldsymbol{z}'}{1 + \boldsymbol{v}_1 \bcdot \boldsymbol{z}_1}\right) \boldsymbol{z}_1,
\end{gather}
where $\boldsymbol{y}_1$, $\boldsymbol{z}_1$, and $\boldsymbol{z}'$ are solutions to
\begin{equation}
    \mathsfbi{A}\boldsymbol{y}_1 = \boldsymbol{r}, \ \ \ \mathsfbi{A}\boldsymbol{z}_1 = \boldsymbol{u}_1, \ \ \ \mathsfbi{A}\boldsymbol{z}' = \boldsymbol{u}_2.
\end{equation}

\section{Implicit Treatment of the Magnetic and Curvature Drift Terms}\label{implicit drifts examples}

When expanded in terms of the Hermite-Laguerre basis functions, the magnetic and curvature drift terms are
\begin{gather}
     \mathrm{i}\frac{\tau_s}{Z_s} \omega_{d}^{\bnabla B} \Big[  (\ell+1) \, H^s_{\ell+1,m} + (2\ell+1)H^s_{\ell,m} +\ell \, H^s_{\ell-1,m} \Big],
\end{gather}
and
\begin{gather}
     \mathrm{i}\frac{\tau_s}{Z_s} \omega_{d}^\kappa \Big[ \sqrt{(m+1)(m+2)} \, H^s_{\ell,m+2} + (2m+1)H^s_{\ell,m} +\sqrt{m (m-1)} \, H^s_{\ell,m-2} \Big].
\end{gather}
The matrix corresponding to this term can be written as
\begin{equation}
    \mathsfbi{B}_{\mathrm{drifts}} = \mathsfbi{B}_{\mathrm{drifts,band}} + \boldsymbol{u}_{\Phi} \otimes \boldsymbol{v}_{\Phi} + \boldsymbol{u}_{A_{\parallel}} \otimes \boldsymbol{v}_{A_{\parallel}},
\end{equation}
where $\mathsfbi{B}_{\mathrm{drifts,band}}$ is a banded matrix of bandwidth $M$. The field terms are contained in the vectors $\boldsymbol{u}_{\Phi}$, $\boldsymbol{v}_{\Phi}$, $\boldsymbol{u}_{A_{\parallel}}$, and $\boldsymbol{v}_{A_{\parallel}}$, whose entries are defined as
\begin{gather}
    \begin{split}
    (u_{\Phi})_{s,\ell,m} &= \mathrm{i} \omega_d^{\bnabla B}\delta_{m0}\left[(\ell+1)\mathcal{J}_{\ell+1} + (2\ell+1)\mathcal{J}_{\ell} + \ell\mathcal{J}_{\ell-1}\right] \\
    &\quad + \mathrm{i}\omega_{d}^\kappa\left[(2m+1)\mathcal{J}_{\ell}\delta_{m0}\right],
    \end{split}\\
    (v_{\Phi})_{s,\ell,m} = \frac{Z_sn_s}{\sum_s \frac{ Z_s^2n_s}{\tau_s}\left(1-\sum_{\ell=0}^{\color{black} N_\ell}(\mathcal{J}^s_\ell)^2\right)}\mathcal{J}_{\ell}^s\delta_{m0},\\
    \begin{split}
    (u_{A_{\parallel}})_{s,\ell,m} &= -\mathrm{i} \omega_d^{\bnabla B}v_{ts}\delta_{m1}\left[(\ell+1)\mathcal{J}_{\ell+1} + (2\ell+1)\mathcal{J}_{\ell} + \ell\mathcal{J}_{\ell-1}\right] \\
    &\quad - \mathrm{i}\omega_{d}^\kappa v_{te}\left[(2m+1)\mathcal{J}_{\ell}\delta_{m1} + \sqrt{m(m-2)}\mathcal{J}_{\ell}\delta_{m3}\right],
    \end{split}
\end{gather}
and
\begin{gather}
     (v_{A_{\parallel}})_{s,\ell,m} = \frac{\beta_{\mathrm{ref}}}{2 \left(k_\perp^2 + \frac{\beta_{\mathrm{ref}}}{2}\sum_s \frac{Z_s^2 n_s}{m_s} \sum_{\ell=0}^{\color{black} N_\ell} (\mathcal{J}^s_\ell)^2\right)}Z_{s}n_sv_{ts}\mathcal{J}_{\ell}^s\delta_{m1}.
\end{gather}
Since this term is evaluated in real space in $z$, $\mathsfbi{B}_{\mathrm{drifts}}$ can be combined with the mirror matrix in the implicit methods.

\begin{figure}
    \centering
    \includegraphics[width=0.5\linewidth]{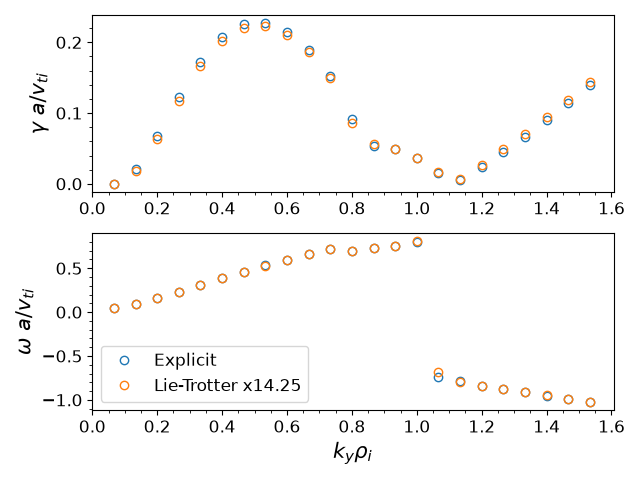}
    \caption{The linear growth rates and frequencies for the explicit RK3 (blue) method and the Lie-Trotter method (orange) scheme for the electrostatic cyclone base-case. The magnetic and curvature drift terms are treated implicitly in this case.}
    \label{fig:implicit drifts}
\end{figure}

To test the implicit treatment of the magnetic drifts, we again run the linear electrostatic CBC simulation. We use the Lie-Trotter method, but the split for the mirror term now also contains the drift terms. For both the mirror and drift terms, both ions and electrons are treated implicitly, and we retain full FLR-effects in the implicit operator. A plot of the linear growth rates and frequencies is in figure \ref{fig:implicit drifts}. In this case, we are able to double the timestep while maintaining good accuracy. The maximum stable timestep is larger, but at the cost of further reduced accuracy. We do not show results using the ARK method, as we find that the operator-split preconditioner requires an increased number of iterations at a larger timestep to remain stable, which offsets the advantage of the increased timestep.

% \begin{gather}
%     \boldsymbol{u}_{\Phi} = \left[-\frac{Z_s}{\tau_s}\mathcal{J}_1^s\Phi -\frac{2Z_s}{\tau_s}\mathcal{J}_0^s\Phi  \quad \boldsymbol{0} \quad -\Phi\delta_{m0}\frac{Z_s}{\tau_s}\left((\ell+1)\frac{}{}\right)\right]
% \end{gather}

% \begin{gather}
%     \boldsymbol{u}_{\Phi} = \left[i\frac{\tau_s}{Z_s}\omega_d^{\kappa}\left(\frac{Z_s}{\tau_s}\mathcal{J}_{\ell}^s\right) + i\frac{\tau_s}{Z_s}\omega_d^{\bnabla B}\left((\ell+1)\mathcal{J}_{\ell+1}\frac{Z_s}{\tau_s} + (2\ell+1)\mathcal{J}_{\ell}\frac{Z_s}{\tau_s} + (\ell)\mathcal{J}_{\ell-1}\frac{Z_s}{\tau_s}\right)\right]
% \end{gather}

\bibliographystyle{jpp}

\bibliography{references}

\end{document}